\documentclass[twocolumn]{aastex7}

\usepackage{xcolor}

\newcommand\ucol{cm$^{-2}$}
\newcommand\ulumi{erg s$^{-1}$}
\newcommand\uflux{erg cm$^{-2}$ s$^{-1}$}

\shorttitle{Quasi-periodic variation in MAXI J0709$-$159}
\shortauthors{Sugizaki et al.}

\graphicspath{{./}{figures/}}

\begin{document}

\title{Quasi-periodic variation during a fast X-ray outburst of a high-mass X-ray binary \\ MAXI J0709$-$159 / LY CMa
  observed by NICER}

\author[0000-0002-1190-0720]{Mutsumi Sugizaki}
\affiliation{National Astronomical Observatories, Chinese Academy of Sciences, 20A Datun Road, Beijing 100012, People's Republic of China}
\affiliation{
Advanced Research Center for Space Science and Technology, 
Kanazawa University, Kakuma, Kanazawa, Ishikawa, 920-1192, Japan}
\email[show]{sugizaki@se.kanazawa-u.ac.jp}

\author[0000-0002-6337-7943]{Tatehiro Mihara}
\affiliation{RIKEN, Hirosawa, Wako, Saitama, 351-0198, Japan}
\email{tmihara@riken.jp}

\author{Kohei Kobayashi}
\affiliation{Department of Physics, Nihon University, 1-8 Kanda Surugadai, Chiyoda-ku, Tokyo, 101-8308, Japan}
\email{csku21001@g.nihon-u.ac.jp}

\author[0000-0001-8195-6546]{Megumi Shidatsu}
\affiliation{Department of Physics, Ehime University, 2-5, Bunkyocho, Matsuyama, Ehime 790-8577, Japan}
\email{shidatsu.megumi.wr@ehime-u.ac.jp}

\author[0000-0002-0207-9010]{Wataru Iwakiri}
\affiliation{International Center for Hadron Astrophysics, Chiba University, Chiba 263-8522, Japan}
\email{iwakiri@hepburn.s.chiba-u.ac.jp}

\author[0000-0001-7115-2819]{Keith Gendreau}
\affiliation{Astrophysics Science Division, NASA Goddard Space Flight Center, Greenbelt, MD 20771, USA}
\email{keith.c.gendreau@nasa.gov}

\author{Zaven Arzoumanian}
\affiliation{Astrophysics Science Division, NASA Goddard Space Flight Center, Greenbelt, MD 20771, USA}
\email{zaven.arzoumanian-1@nasa.gov}

\author{Douglas J. K. Buisson}
\affiliation{Independent}
\email{djkbuisson@gmail.com}

\author[0000-0002-8403-0041]{Sean N. Pike}
\affiliation{Center for Astrophysics and Space Sciences, University of California, San Diego, CA 92093}
\email{snpike@ucsd.edu}


\begin{abstract}
We report on a quasi-periodic variation at $\sim1$ Hz during a fast X-ray
outburst of a high-mass X-ray binary MAXI J0709$-$159 / LY CMa observed by the Neutron-star interior composition explorer (NICER).
The new X-ray transient MAXI J0709$-$159 was discovered on 2022 January 25.
Due to the transient X-ray behavior characterized by the short (a few hours) 
outburst duration, rapid ($\lesssim$ 1 s) variability with spectral
change, and large luminosity swing 
from  $10^{32}$ {\ulumi} 
to $10^{37}$ {\ulumi}, 
the object was considered likely to be a supergiant X-ray binary with a neutron star (NS) categorized as a 
Supergiant Fast X-ray Transient (SFXT).
Follow-up NICER and NuSTAR observations confirmed that the position of the new X-ray object  
is consistent with a Be star, LY CMa, 
which has been also identified as a B supergiant.
We analyzed the NICER data 
obtained from 3 hours to 6 days after the discovery.
The light curve reveals that the X-ray activity continued for $\sim7$ hours
in sparse short flares,
each lasting $\lesssim 100$ seconds, 
and the luminosity instantaneously reached up to $\sim 1\times 10^{38}$ {\ulumi}.
The light-curve and spectral features
reasonably agree with those expected from accretion of a clumpy stellar-wind  
onto a magnetized NS.
The variability power spectrum during the brightest flare
shows a broad peak at $1.1$ Hz
resembling a quasi-periodic oscillation (QPO).
If the QPO is attributed to the Keplerian orbital frequency at the inner edge
of a transient accretion disk truncated by the NS magnetosphere,
the NS surface magnetic field is estimated to be $\sim 10^{12}$ G.
\end{abstract}


\section{Introduction} \label{sec:intro}

MAXI J0709$-$159 \citep[hereafter MAXI J0709;][]{serino2022ATel} is a
new X-ray transient discovered on 2022 January 25
near the Galactic plane at
$(l,b)=(229\fdg3, -2\fdg3)$
by the Monitor of All-sky X-ray Image 
\citep[MAXI; ][]{matsuoka09}
onboard the International Space Station
(ISS).
From the observed transient behavior characterized 
by the short ($\lesssim 3$ hours) activity duration, 
rapid ($\lesssim$ a few seconds) variability
accompanied with spectral changes, and large
luminosity swing by a factor of $\gtrsim 10^{4}$ from the quiescence to the outburst peak,
the object was considered likely to be a Supergiant Fast X-ray Transient (SFXT),
a possible subclass of 
supergiant X-ray binaries (SgXBs) 
accompanied by magnetized neutron stars (NSs) \citep{kobayashi2022ATel}.

By coordinated follow-up observations with the Neutron star Interior Composition
ExploreR \citep[NICER;][]{2016SPIE.9905E..1HG} and
the Nuclear Spectroscopic Telescope Array \citep[NuSTAR;][]{harrison2013},
MAXI J0709 was successfully identified with a new X-ray object 
located at the position consistent with an optical companion, LY CMa,
also known as HD 54786
\citep{iwakiri2022ATel, negoro2022ATel},
located at the distance of $D=3.03^{+0.31}_{-0.27}$ kpc \citep{2021AJ....161..147B}.
%
LY CMa has been categorized as a Be star due to the presence of hydrogen emission lines 
\citep{2015AJ....149....7C}
but is also known as a B supergiant (B1.5b) in the optical spectral classification \citep{1988mcts.book.....H}.
Hence, optical follow-up observations were carried out. 
The results confirmed that the optical spectra had a broad H$\alpha$ emission line
suggesting that the circumstellar Be disk was formed 
\citep[][]{nesci2022ATel,bhattacharyya2022,sugizaki2022,2025arXiv250321118S}.
These observed X-ray and optical features suggest
that the optical counterpart, LY CMa, certainly
has a complex circumstellar medium (CSM)
embracing the Be disk as well as 
clumpy stellar winds \citep{sugizaki2022}.
Also, \citet{2024BSRSL..93..636B} proposed   
that the object is on an intermediate evolution phase between Be X-ray binaries (BeXBs) and SgXBs 
from its optical color-magnitude location.

SFXTs are proposed as a possible subclass of SgXBs for those members
that sporadically  (once every few months to several years) exhibit 
short-duration (several hours) outbursts
(e.g. \citealt{2006ApJ...646..452S, 2015AdSpR..55.1255B,2013AdSpR..52.1593R, 2018MNRAS.481.2779S};
review in \citealt{2019NewAR..8601546K}).
So far, about a dozen SFXTs and their candidates have been identified in our Galaxy.
Because a few of them were found to show coherent X-ray pulsations,
these objects are considered likely to be binary systems hosting magnetized NSs.
However, physical mechanisms that enable such extremely short outbursts
have not been well understood.
These features in X-ray activity 
are quite different from those of classical SgXBs
that usually exhibit persistent X-ray emission.
From the sporadic and short-duration outburst activities, 
mass accretion onto the NS is thought to be
induced by interaction with clumpy stellar winds.
The large intensity swing by a factor of $\sim 10^{5}$
from the quiescence to the outburst peak in the short time period
requires
some mechanisms
to inhibit mass accretion such as magnetic and/or centrifugal
barriers \citep{2007AstL...33..149G, 2008ApJ...683.1031B}.
To explain this, 
\citet{2008ApJ...683.1031B} proposed a hypothesis that the NS should have a strong
surface magnetic field $B_\mathrm{s} \sim 10^{13}\mathord-10^{14}$ G like a magnetar.
On the other hand, \citet{2013MNRAS.428..670S,2014MNRAS.442.2325S} 
proposed another scenario that these objects are slow-rotating NSs with regular $B_\mathrm{s}$ ($\sim 10^{12}$ G)
and the sporadic outbursts result from instabilities of quasi-spherical accretion shells
formed around the NS magnetosphere.

MAXI J0709 is a unique high-mass X-ray binary (HMXB)
that exhibited transient behavior like SFXTs
but has a stellar companion which shows properties of Be stars as well as B supergiants.
Hence, it can become a critical sample to address questions regarding SFXTs.
Usually, observations of SFXTs in their active outburst phase
are difficult because their occurrence is very rare and unexpected.
The NICER observations of MAXI J0709,
which started just 3 hours after the discovery,  
were enabled by
the coordinated operation between MAXI and NICER,
two astronomical experiments on the ISS.
The NICER data provides us a great opportunity to study the rapid variability
with large photon statistics and excellent time resolution
that had not been realized before.

In this paper, we present detailed analysis of
the NICER observations of MAXI J0709.
We describe the NICER observations
in section \ref{sec:observation},
and the data analysis and results in section \ref{sec:analysis}.
Based on the results, 
we discuss possible mechanisms for the observed fast X-ray variability 
in section \ref{sec:discussion}.
Throughout the paper below, we provide errors on observed parameters
at 90\% confidence limits of statistical uncertainties,
unless otherwise specified.

\section{Observations and data reduction}
\label{sec:observation}

NICER is an X-ray astronomical mission
that has been carried out on the ISS since 2017. 
The main science instrument, X-ray Timing Instrument (XTI),
consists of 56 X-ray concentrator optics (XRC)
and matching silicon drift detectors, 
covering the 0.2--12 keV band with large effective area (1700 cm$^{2}$ at 1 keV).
The detectors have a capability of recording every X-ray event
with unprecedented time resolution ($\lesssim 300$ ns)
and energy resolution similar to standard silicon-based semiconductor detectors (137 eV at 6 keV)
\citep{2016SPIE.9905E..1HG}.

The first NICER observation of MAXI J0709 was performed to
identify the new X-ray transient detected
by MAXI on 2022 January 25 
under the MAXI-NICER coordinated observation program.
At UT 13:42, approximately 3 hours after the first MAXI detection, 
the NICER/XTI grid scan observation covering the entire MAXI error region
encircled by a radius of $\sim 0\fdg25$ started.
At UT 13:54,
the new X-ray object was successfully confirmed
and the source position was determined
with $3\arcmin$ accuracy (appendix \ref{sec:gridscan}).
Subsequently, additional follow-up observations to monitor the source activity
were carried out.
Table \ref{tab:obslog} lists all of the NICER observations 
performed in the five days after the discovery.
The source position was later refined 
with $10\arcsec$ accuracy
by the NuSTAR observation
and then the optical counterpart of LY CMa was identified
\citep[][]{sugizaki2022}.

We analyzed all NICER data including grid-scan data
listed in table \ref{tab:obslog}.
The data reduction and analysis
were performed with the standard analysis tools NICERDAS version v011a released as a part of HEASOFT version 6.32.1
and the calibration files CALDB version xti20221001.
We first reprocessed all event data with a standard tool {\tt nicerl2}
and screened them with typical filtering criteria:
the pointing offset is less than $3\arcmin$ of the source position uncertainly (appendix \ref{sec:gridscan}),
the pointing direction is more than $40\degr$ away from the bright Earth limb,
more than $30\degr$ away from the dark Earth limb,
the ISS is located outside the South Atlantic Anomaly (SAA),
and the geomagnetic cutoff rigidity (COR) is $>1.5$ GeV$/c$.
Events measured by two noisy detectors, FPMs 34 and 14,
were also filtered out.
We started 
all data analysis in the following sections
with the obtained cleaned data.

\begin{deluxetable}{lccr}
  \tablecaption{Logs of NICER MAXI J0709$-$159 observations performed in 2022 January.
  \label{tab:obslog}}
\tablewidth{0pt}
\tablehead{
  \colhead{ObsID} & \colhead{Start Time} & \colhead{Stop Time} & \colhead{Exp.}\\
  \colhead{} & \colhead{(UTC)} & \colhead{(UTC)} & \colhead{(s)} 
}
\startdata
4202520101\tablenotemark{\small a} & Jan 25 13:54:11 & Jan 25 23:27:07 & 7174 \\
4202520102 & Jan 26 00:35:34 & Jan 26 22:39:46 & 12728 \\
4202520103 & Jan 26 23:49:13 & Jan 27 03:21:20 & 3519 \\
4202520104 & Jan 28 00:35:53 & Jan 28 00:49:02 & 789 \\
4202520105 & Jan 29 02:55:54 & Jan 29 11:06:40 & 4830 \\
4202520106 & Jan 30 08:27:34 & Jan 30 16:26:23 & 2016 \\
\enddata
\tablenotetext{\small a}{For the first 30 minutes, a grid scan observation was done.}
\end{deluxetable}

\section{Analysis} \label{sec:analysis}

\subsection{Overall light curve}
\label{sec:lcall}

\begin{figure*}
\begin{center}
\includegraphics[width=0.85\textwidth]{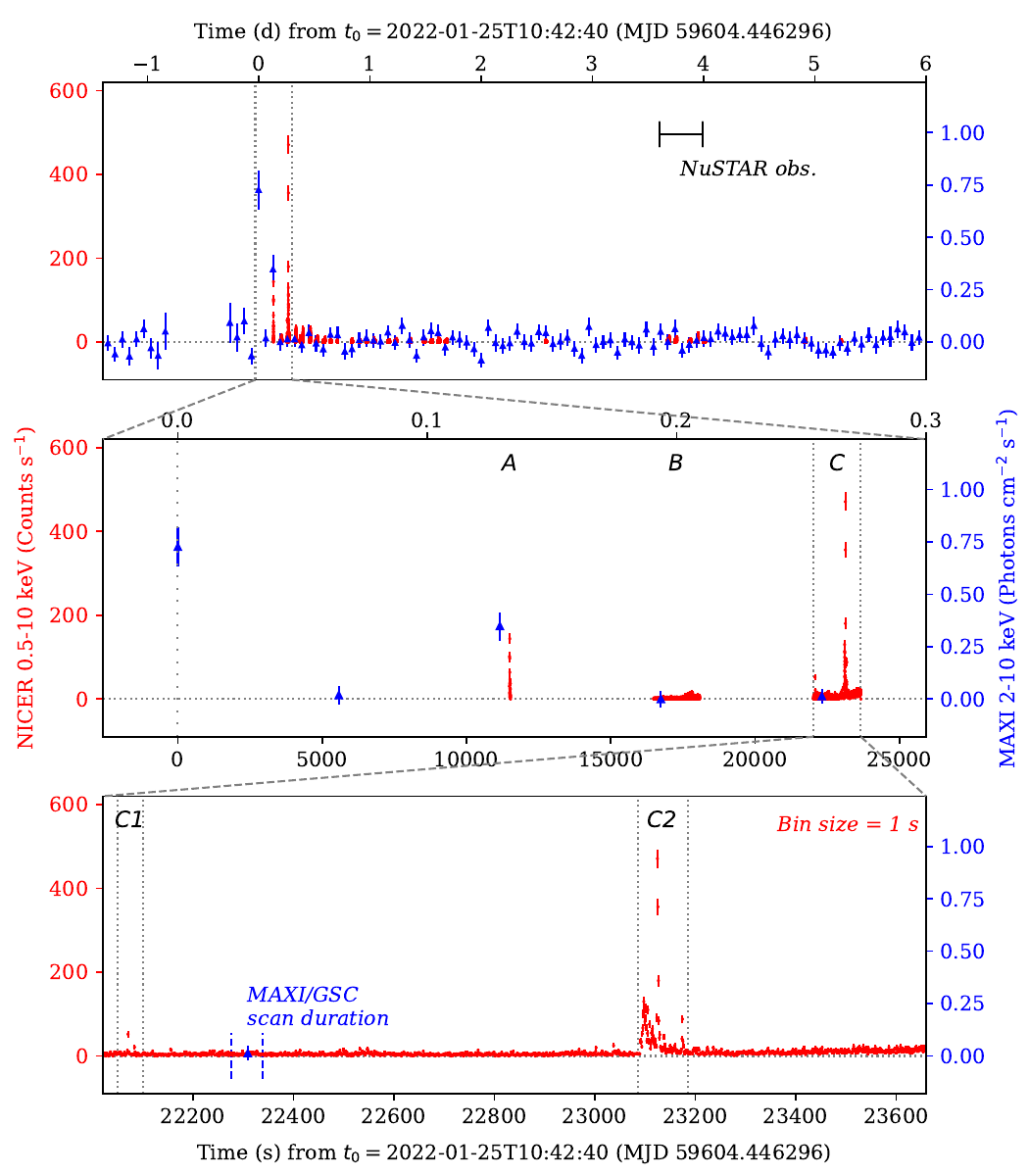}
\caption{
MAXI J0709$-$159 X-ray light curves
observed by NICER in 0.5--10 keV band (red)
and MAXI in 2--10 keV band (blue).
The NICER data represents count rate including backgrounds
every 1-s time bin.
MAXI data are obtained from the public web site, which
provides estimated photon flux every scan transit of $\sim 40$ s
assuming the power-law spectrum of photon index $\Gamma=2$.
X-axis labels at the top and bottom on each panel represent the elapse
time from the first source detection at $t_0=$ UTC 2022 January 25 10:42:40
(MJD 59604.446296) in units of day and second, respectively.
%
%
{\bf (Top)} Overall profiles of 7 days including all NICER observations
from one day before the first MAXI detection, annotated with the NuSTAR observation period.
{\bf (Middle)} First 6 hours including all significant X-ray flares observed by NICER.
Labels A, B, C represents three good-time intervals of NICER data separated by the 90-min ISS orbital period.
{\bf (Bottom)} 20 minutes of the NICER GTI-C including the brightest flare in the NICER data.
Labels C1, C2 represent periods of two remarkable flares.
}
\label{fig:lcmxnr}
\end{center}
\end{figure*}

To overview the entire outburst profile, 
we first extracted light curves of all NICER data
and plotted them with the MAXI/GSC
\citep[Gas Slit Camera;][]{2011PASJ...63S.623M,2011PASJ...63S.635S}
data, which
obtained from the MAXI quick-look website\footnote{http://maxi.riken.jp}.
Figure \ref{fig:lcmxnr} (top) shows the obtained NICER 0.5--10
keV light curve with 1-s time bins
and the MAXI 2--10 keV data taken every 90-min ISS orbital cycle
for 7 days from 1 day before the first MAXI detection.
There, the scales of X-ray intensities
measured by two instruments, NICER/XTI and MAXI/GSC,
are normalized by assuming the typical power-law spectral model with
a photon index $\Gamma=1$
and an absorption column density $N_\mathrm{H}=10^{23}$ {\ucol},
which were determined in the spectral analysis (section \ref{sec:ana_spec}).
In this power-law model,
the NICER peak count rate, $\sim 500$ cts s$^{-1}$,
corresponds to the 0.5--10 keV flux of $\sim 1\times 10^{-8}$ {\uflux}.
%
Hereafter,
the time from the first MAXI source detection at $t_0=$ 2022-01-25 10:42:40 (UTC)
is denoted as $t$.

The combined NICER and MAXI light curves reveal that
the source exhibited several flares only
in the initial $0.3$ days $\simeq 7$ hours. 
After $t=0.3$ d, 
no significant X-rays from the target source
have been detected with either NICER or MAXI.
At $t=3.5$ d, a NuSTAR observation of 18 ks exposure was carried out
and the new X-ray object with the 2--10 keV flux of $6\times 10^{-13}$ {\uflux}
was detected \citep{sugizaki2022}.  
This means that
the source intensity decreased by a factor of $10^{-4}$ in 3.5 days.
The time interval of the NuSTAR observation is
illustrated in the figure \ref{fig:lcmxnr} (top).

Figure \ref{fig:lcmxnr} (middle) shows
the close-up view of the initial 0.3 days.
Because both MAXI and NICER are ISS-payload missions,
the good-time intervals (GTIs) of these observations 
are synchronized with the ISS orbital period 
of $\sim 90$ minutes ($=5400$ s).
Hence, we named the three GTIs of the NICER data
in this 0.3 days 
as GTI-A, B, C, as labeled in the figure.
As seen in the figure,
data taken at the same time or close in time
by MAXI and NICER are roughly consistent.

The observation of the GTI-A was performed 
as a part of the grid scan to identify the new transient. 
The position of the later identified object, LY CMa,
was observed only for $37$ s from $t= 11491$ s to $11528$ s.
The NICER light curves (red dot marks) shows a remarkable X-ray flare
with a large variability.
At $t=11150$ s, i.e. 340 s before the GTI-A,
MAXI/GSC scanned the source position
and detected a significant flare
with the same activity level (blue triangle marks).
This suggests that this flare would 
continue over the two observations separated by 340 s.
In the next GTI-B covering $1592$ s from $t=16488$ s to
$18080$ s,
no significant X-rays over the backgrounds were
detected.
However, in the GTI-C of $1640$ s from $t=22020$ s to $23660$ s,  
bright X-ray flares were detected again for a part of the interval.
The peak intensity in the GTI-C was highest among all of the NICER data
and comparable
to that at the initial source onset ($t=0$) observed by MAXI.

Figure \ref{fig:lcmxnr} (bottom) shows the further close-up view
of the GTI-C. 
The flaring activities appeared only in a few short intervals
of $\lesssim100$ s.
The brightest flare at around $t=23100$ s
exhibits rapid variations in time scales 
shorter than the 1-s bin size.
Hereafter,
we refer to the two clear flaring periods 
labeled in the figure
as C1 and C2.

\subsection{Rapid variability and hardness ratio} 
\label{sec:lcfine}
From the overall light curve of figure \ref{fig:lcmxnr}, 
we found that the X-ray activity of 
MAXI J0709
appeared only until $t=0.3$ d ($\simeq 7$ h) 
with several short flare-up episodes, 
each lasting only $\lesssim 100$ s.
Hence, we next investigated the details of each flare.

\begin{figure*}
\begin{center}
\includegraphics[width=1.\textwidth]{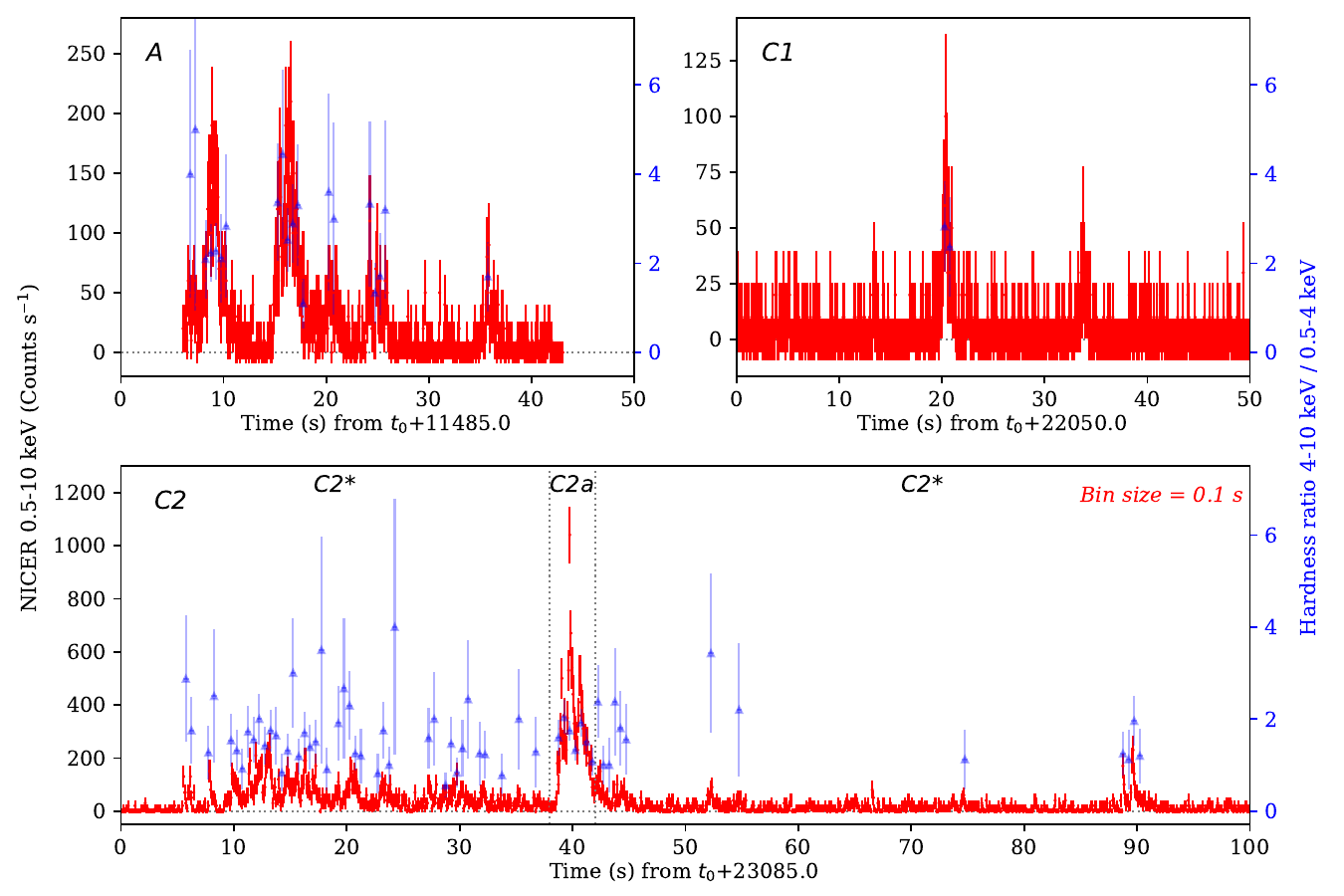}
\caption{
NICER 0.5--10 keV light curve in 0.1-s time bin (red)
and hardness ratio variation of 4--10 keV to 0.5--4 keV in 0.5-s time bin (blue)
for three flaring periods of A (50 s), C1 (50 s), and C2 (100 s), labeled in figure \ref{fig:lcmxnr}.
}
\label{fig:lcmulti}
\end{center}
\end{figure*}

Figure \ref{fig:lcmulti} shows the light curves 
of three remarkable flaring periods,
GTI-A (hereafter simply A), C1 and C2,
labeled in figure \ref{fig:lcmxnr},
for 50 s, 50 s and 100 s, respectively,
with 0.1-s time bins.
These profiles reveal that each flare consists of 
short spikes of a few seconds with a similar triangular shape.
The flare C2 includes the brightest interval of a few seconds
with the count rate over $200$ cts s$^{-1}$,
which is significantly higher than that in the rest of the period.
We named this brightest 4 seconds as C2a, as labeled in figure \ref{fig:lcmulti},
and the remaining C2 except C2a as C2*.
The peak count rate of $\sim 1000$ cts s$^{-1}$ in the 0.1-s time bin is twice as high as 
$\sim 500$ cts s$^{-1}$ in the 1-s time bin (figure \ref{fig:lcmxnr}),
indicating that the variation time scale is shorter than 1 s.

\begin{figure}
\begin{center}
\includegraphics[width=0.44\textwidth]{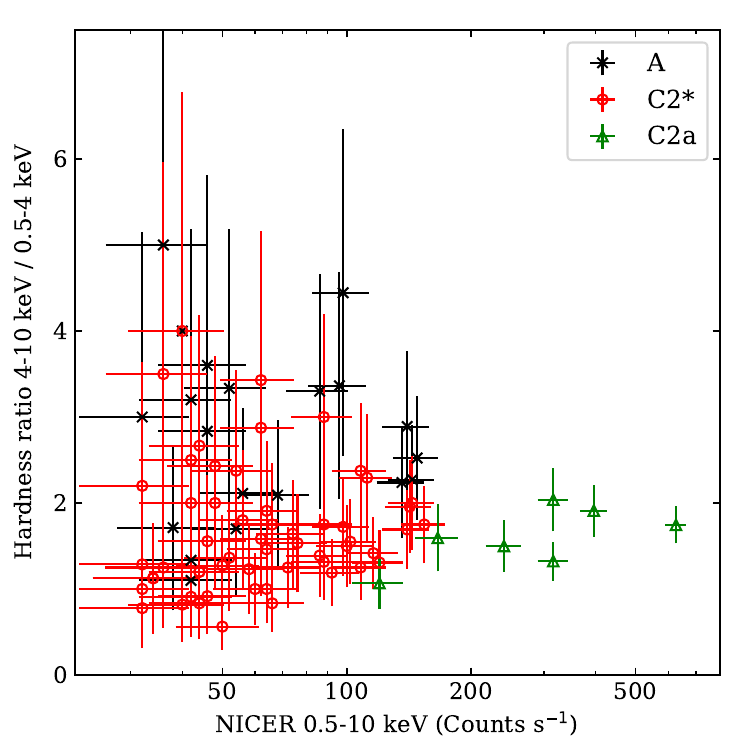}
\caption{
  Hardness-intensity diagram during the two flares of
  A and C2, shown in figure \ref{fig:lcmulti}.
}
\label{fig:hid}
\end{center}
\end{figure}

To investigate the spectral change synchronized with the intensity variation,
we plot the 4--10 keV to 0.5--10 keV hardness ratio of every 0.5-s time bin
on the light curves in figure \ref{fig:lcmulti} (with blue dot).
Also, the hardness-intensity diagram is shown in figure \ref{fig:hid},
where data of three distinctive flaring periods, A, C2a and C2*,
are marked with different colors and symbols.
We cannot see any clear correlation between the hardness and intensity.
On the other hand,
the hardness of the flare A is slightly higher than that of the flare C2
(including both C2a and C2*).

\subsection{Variability power spectrum}

To study time scales of the fast variability seen in figure \ref{fig:lcmulti}
and also search for possible coherent pulsations,
we investigated the power-density spectra (PDS).
In the beginning, we converted all event times, 
which were recorded on the ISS, 
into the arrival times at the solar system barycenter 
using {\tt barycorr}, a standard tool in HEASOFT, with the ISS orbit data
and assuming
that all X-ray events came from the direction of the identified object, LY CMa.
We used the Python package {\tt stingray} \citep{2019ApJ...881...39H}
to calculate the PDS from the event data.

We analyzed data of three GTIs, A, B, and C, defined
in section \ref{sec:lcall} as shown in figure \ref{fig:lcmxnr} (middle), separately.
Because the GTI-B shows no X-ray activity, 
the result can be used as the reference for background data.
We first calculated PDS for every time segment divided by a specific length
from light-curve data in 0.5-ms time bin, 
i.e. in the frequency range up to 1 kHz.
The lower end of the frequency range depends on the segment length. 
Considering the length of each GTI, 
we divided the GTI-A of 37 s into two 18.5-s segemets,
the GIT-B of 1592 s into four 398-s segments,
and the GTI-C of 1640 s into four 410-s segments.
Then, we calculated the average PDS of each GTI.

\begin{figure*}
\begin{center}
\includegraphics[width=1.\textwidth]{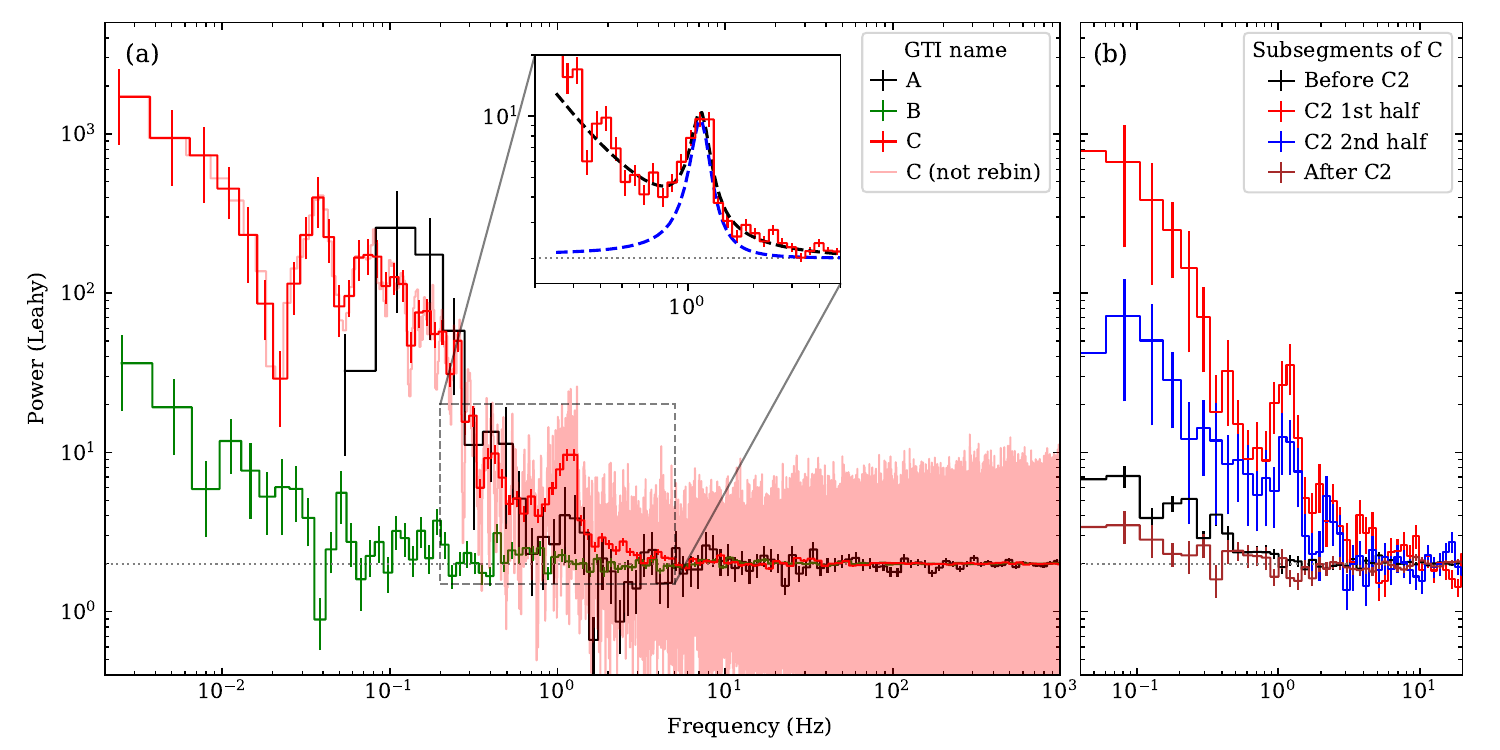}
\caption{
  (a) Power density spectra (PDS) of NICER 0.5--10 keV light curves
  in three GTIs, A, B, and C, labeled in 
  figure \ref{fig:lcmxnr} (middle).
  The inset shows the close-up view of the peak seen in the PDS of GTI-C
  at $\sim1$ Hz, where dashed lines represent the best-fit 
  Lorentzian plus power-law model in black and 
  its Lorentzian component in blue.
  (b) PDS of four GTI-C subsegments of
  before C2 flare,  first half and second half of C2, and after C2. 
}
\label{fig:pds}
\end{center}
\end{figure*}

Figure \ref{fig:pds} (a) shows the obtained PDSs of the three GTIs
with Leahy normalization \citep{1983ApJ...266..160L} and logarithmical rebinning for visual clarity.
As a representative example, an unbinned PDS of the GTI-C is shown together.
No significant signal of coherent pulsations was detected.
%
The PDSs of GTI-A and GTI-C, which have significant
X-ray activities, show a similar profile over the background estimated
from the GTI-B data.
The broad enhanced structure below 0.5 Hz corresponds to
a cycle of short spikes every $\sim 10$ s.
In addition, an isolated peak at $\sim 1$ Hz with a certain width
like quasi-periodic oscillation (QPO)
is clearly seen in the GTI-C.
In the higher frequency above $\sim1$ Hz,
all PDSs are consistent with a constant $=2$
corresponding to the Poisson fluctuation
in the Leahy normalization.

To examine possible time variation of the QPO-like feature during the GTI-C,
we calculated PDSs for 4 divided subsegments of
(i) before the bright-flare period C2,
(ii) the first half of C2,
(iii) the second half of C2 including the flare peak, and 
(iv) after C2.
Figure \ref{fig:pds} (b) shows the obtained PDSs. 
The QPO-like signal can be recognized
in both first half and second half of C2
at the same frequency $\sim 1$ Hz, but not anywhere else.
Therefore, the feature is considered to associate with the C2 flare.

To characterize the QPO-like feature,
we fitted the PDS profile around the peak 
from 0.25 Hz to 5 Hz
with a model consisting of a Lorentzian function,
a power-law continuum, and a constant offset $=2$ for the Poisson fluctuation,
which has been often used for QPOs in various X-ray binaries 
\citep[e.g.][]{2002ApJ...572..392B}.
The fit is accepted with the best-fit Lorentzian parameters of
the centroid frequency $\nu_0=1.14\pm 0.03$ Hz,
half-width at the half-maximum $\Delta=0.12\pm0.02$,
integrated fractional RMS $=0.58\pm0.05$,
and the continuum power-law index of
$-1.59_{-0.20}^{+0.16}$.
The quality factor for the QPO feature is estimated as $Q=\nu_0/2\Delta=4.8$.
In the inset of figure \ref{fig:pds},
the best-fit model is shown on the data.

\subsection{Energy spectra}
\label{sec:ana_spec}

We investigated X-ray energy spectra of individual flare periods.
We used the standard NICER analysis script {\tt nicerl3-spect}
to extract source spectra, background spectra predicted by
SCORPEON model\footnote{https://heasarc.gsfc.nasa.gov/docs/nicer/analysis\_threads/scorpeon-overview/},
and energy response functions for specified GTIs.
We performed model fitting on XSPEC version 12.13.1 \citep{arn96}
based on the $W$-statistic, a modified version of C-statistic \citep{Cash1979}
taking into account the background contributions.
There, 
using the {\tt error} command 
or 
the {\tt eqwid} command (for equivalent width) 
in XSPEC, 
we calculated model-parameter errors
corresponding to an increase of 2.7 in the fit statistic $W$,
which are equivalent to 90\% confidence limits.

\begin{figure*}
\begin{center}
\includegraphics[width=1.\textwidth]{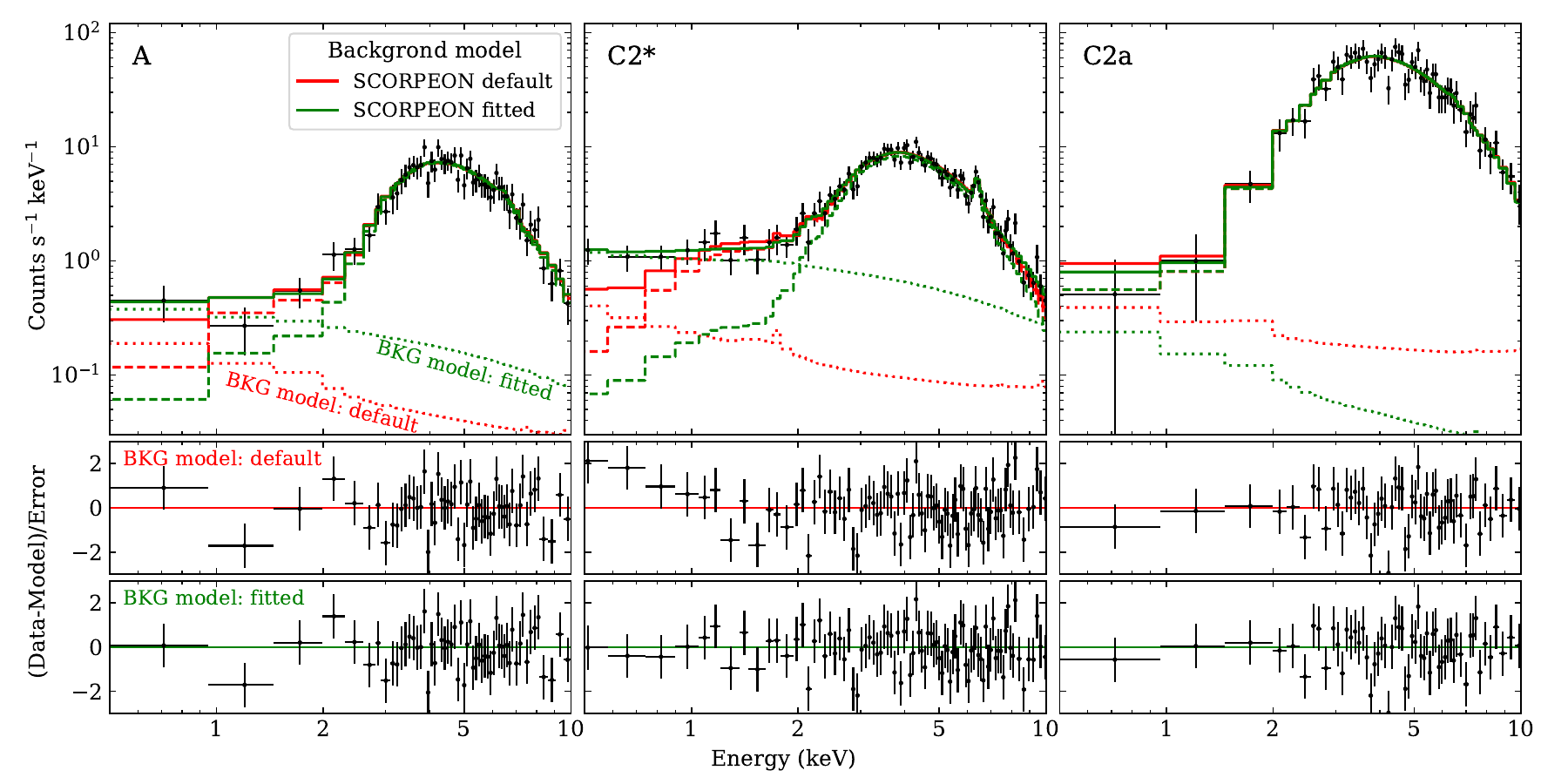}
\caption{
  X-ray spectra of three flaring periods,
  A, C2* (C2 excluding C2a), and C2a, labeled in figure \ref{fig:lcmulti}.
  {\bf (Top)}  Observed photon count spectra including backgrounds
  and best-fit models folded with instrument response functions (solid lines).
  Dashed and dotted lines represent 
  partial-covering-absorption power-law plus iron-K$\alpha$ line models for source spectra
  and background models with SCORPEPN default (red) and those fitted (green).
  {\bf (Middle and bottom)}
  Data-to-model residuals with default SCORPEON background models and those
  allowed to fit in the source model fitting.
}
\label{fig:espec}
\end{center}
\end{figure*}

Figure \ref{fig:espec} shows the obtained energy spectra
including backgrounds
for three distinctive flaring periods, A, C2* (C2 except C2a peak) and C2a,
which have been defined in section \ref{sec:lcfine} as in figure \ref{fig:lcmulti}.
We first fitted all the spectra with a power-law model 
with interstellar-medium (ISM) absorption
employing the Tuebingen-Boulder absorption model ({\tt tbabs})
with the solar abundance given by \citet{Wilms2000}
and background models predicted by the default SCORPEON parameters.
As the results, we obtained similar best-fit model parameters of 
power-law photon index $\Gamma\simeq 1$ and 
absorption hydrogen column density $N_\mathrm{H}\simeq 10^{23}$ {\ucol}
for all three spectra.
The obtained $N_\mathrm{H}\simeq 10^{23}$ {\ucol} is significantly higher than the Galactic
hydrogen ($\mathrm{H}_\mathrm{I}$) column density in the source direction, 
$0.55\times 10^{22}$ {\ucol} \citep{hi416}, 
meaning that the primary X-ray emission from the central X-ray object (presumably NS)
is further obscured by local CSM.
The goodness of the fits
represented by $W$ for degrees of freedom (d.o.f.)
are not within the acceptable level 
for the two spectra of A and C2*.
The data-to-model residuals show excess remaining 
in the lower energy band below 1 keV
and a possible iron-K$\alpha$ line at 6.4 keV.

We hence modified the simple power-law model by multiplying a partial covering-absorption factor ({\tt tbpcf}) for the local CSM
and adding a narrow Gaussian function for the iron-K$\alpha$ line,
and then fit it to the data.
The modified model is expressed by {\tt tbabs*tbpcf*(powerlaw+gauss)} in XSPEC terminology.
There, $N_\mathrm{H}$ for the ISM in the {\tt tbabs} model 
was fixed at the Galactic $\mathrm{H}_\mathrm{I}$ density 
$=0.55\times 10^{22}$ {\ucol},
while the partial-covering absorption parameters
for the CSM in the {\tt tbpcf} model 
were allowed to vary.
The Gaussian centroid and width for the iron-K$\alpha$ line
were fixed at 6.4 keV and 0.1 keV,
those typical values in HMXBs \citep{2015A&A...576A.108G, 2018A&A...610A..50P},
respectively.
Also, the background models were fixed at the default SCORPEON.
Then, the fits became acceptable in all three spectra.
In figure \ref{fig:espec}, the best-fit models folded with the instrument response functions
and the data-to-model residuals are shown.
The uncovered fraction of the partial covering absorber
$1-f_\mathrm{pc}$ is suggested to be significantly $>0$
in A and C2*  
although as small as $\sim10^{-2}$.
The iron-K$\alpha$ line is positively detected
only in the C2* spectrum with a confidence level of $99$\% or higher.

The systematic excess residuals below 1 keV in the first model fits
may be partly due to the errors on the SCORPEON background models.
To examine the impact of the background-model errors, 
we repeated the model fitting by floating the SCORPEON model parameters
within their allowed ranges.
In figure \ref{fig:espec}, 
the best-fit background and emission models
are overlaid on those obtained with the default SCORPEON models.
The difference is clear in the C2* spectrum.
The predicted background level increased by a factor of $\sim5$ 
and the uncovered fraction of the partial covering absorber
$1-f_\mathrm{pc}$ turned to accept 0.
On the other hand, the iron-K$\alpha$ line parameters almost unchanged.

Table \ref{tab:specparams} lists all the best-fit spectral parameters including
power-law photon index $\Gamma$, 
column density of the partial covering absorber $N_\mathrm{H,pc}$,
uncovered fraction $1-f_\mathrm{pc}$,
iron-K$\alpha$ line flux $F_\mathrm{FeK\alpha}$ and equivalent width $EW_\mathrm{FeK\alpha}$,
model flux with ISM/CSM absorption $F_\mathrm{abs}$,
absorption-corrected flux $F_\mathrm{unabs}$,
and absorption-corrected luminosity $L_\mathrm{ubabs}$
assuming the isotropic source emission at the 3 kpc distance.
$F_\mathrm{abs}$, $F_\mathrm{unabs}$, and $L_\mathrm{ubabs}$
are calculated in the observed energy band of 0.5--10 keV.
There, results obtained with two background models of the default SCORPEON model parameters
and those allowed to fit to the data are shown together.

Comparing the obtained model parameters in A 
with those obtained from the MAXI/GSC data 
for a possible continuous flare \citep{sugizaki2022},
$\Gamma\simeq1$ in the present NICER results
is smaller than $\Gamma\simeq2$ in the MAXI data.
This is considered from the difference in the energy band
between the NICER/XTI 0.5--10 keV and the MAXI/GSC 2--20 keV
and suggests that
the power-law model should have a break or cutoff at around 10 keV.
We hence fitted the MAXI/GSC spectrum in \citet{sugizaki2022} with an exponential cutoff power-law
({\tt cutoffpl} in Xspec terminology) 
with $\Gamma=1$ and $N_\mathrm{H}=1.4\times 10^{23}$ {\ucol} fixed
at the present NICER results, 
and obtained the e-folding energy 
$E_\mathrm{fold}=22^{+\infty}_{-13}$ keV.
In this cutoff power-law model,
the conversion factor from the 0.5--10 keV flux to
the bolometric flux is $\sim2.9$.

We also attempt to fit with a model
employing a blackbody continuum in place of the power-law continuum,
represented by {\tt tbabs*tbpcf*(bbodyrad+gauss)} in Xspec terminology.
The obtained $W$ for d.o.f. are not as good as those for
the power-law continuum model but close to the acceptable level.
The best-fit parameters for each of the three spectra are 
shown together in Table \ref{tab:specparams}.
The blackbody temperature $kT_\mathrm{BB}\sim 2$ keV is consistent with
that of the MAXI/GSC results in the same flare period
\citep{sugizaki2022}.

\begin{deluxetable*}{l@{\hspace{1cm}}ccc@{\hspace{1cm}}ccc} 
\tablecaption{
  Best-fit spectral parameters of partial-covering-absorption power-law or blackbody model
  for three flaring periods A, C2* (C2 excluding C2a), and C2a
  with default or fitted SCORPEON background model.
  \label{tab:specparams}
}
\tablehead{
  \multicolumn{1}{c}{Flare ID name} & {A} & {C2*} & {C2a} & {A} & {C2*} & {C2a}\\
  \multicolumn{1}{c}{Background model} & \multicolumn{3}{c}{SCORPEON default} & \multicolumn{3}{c}{SCORPEON fitted} 
}
\startdata
\hline
\multicolumn{1}{c}{Spectral model} & \multicolumn{6}{c}{\tablenotemark{\small a}\tt tbabs*tbpcf*(powerlaw+gauss)}\\
$N_\mathrm{H,pc}$ ($10^{22}$ cm$^{-2}$)\tablenotemark{\small b} & $14.0^{+2.5}_{-2.4}$ & $10.2^{+1.0}_{-1.0}$ & $7.9^{+1.4}_{-1.2}$ & $14.3^{+2.7}_{-2.5}$ & $11.3^{+1.4}_{-1.3}$ & $7.8^{+1.4}_{-1.2}$\\
$1-f_\mathrm{pc}$ ($10^{-2}$) & $0.42^{+0.41}_{-0.24}$ & $1.15^{+0.44}_{-0.33}$ & $<0.32$ & $0.24^{+0.34}_{-0.22}$ & $<0.28$ & $<0.36$\\
$\Gamma$ & $1.07^{+0.36}_{-0.35}$ & $1.32^{+0.20}_{-0.20}$ & $0.83^{+0.30}_{-0.29}$ & $1.12^{+0.38}_{-0.37}$ & $1.62^{+0.26}_{-0.25}$ & $0.80^{+0.30}_{-0.29}$\\
$N_\mathrm{PL}$ (photons cm$^{-2}$ s$^{-1}$ at 1 keV) & $0.11^{+0.11}_{-0.05}$ & $0.14^{+0.06}_{-0.04}$ & $0.44^{+0.32}_{-0.18}$ & $0.12^{+0.13}_{-0.06}$ & $0.22^{+0.14}_{-0.08}$ & $0.42^{+0.30}_{-0.17}$\\
$F_\mathrm{FeK\alpha}$ ($10^{-11}$ {\uflux})\tablenotemark{\small c} & $<2.4$ & $2.2^{+1.1}_{-1.0}$ & $<13.3$ & $<2.5$ & $2.4^{+1.1}_{-1.0}$ & $<13.1$\\
$EW_\mathrm{FeK\alpha}$ (eV)\tablenotemark{\small c} & $<155$ & $173^{+99}_{-70}$ & $<138$ & $<183$ & $212^{+115}_{-85}$ & $<112$\\
$F_\mathrm{abs}$  ($10^{-9}$ {\uflux})\tablenotemark{\small d} & $0.71^{+0.06}_{-0.06}$ & $0.66^{+0.03}_{-0.03}$ & $5.23^{+0.45}_{-0.45}$ & $0.69^{+0.06}_{-0.06}$ & $0.56^{+0.03}_{-0.03}$ & $5.28^{+0.46}_{-0.44}$\\
$F_\mathrm{unabs}$ ($10^{-9}$ {\uflux})\tablenotemark{\small d} & $1.54^{+0.40}_{-0.24}$ & $1.42^{+0.20}_{-0.14}$ & $8.48^{+1.23}_{-0.72}$ & $1.56^{+0.45}_{-0.26}$ & $1.54^{+0.38}_{-0.24}$ & $8.49^{+1.12}_{-0.73}$\\
$L_\mathrm{unabs}$ ($10^{36}$ {\ulumi})\tablenotemark{\small e} & $1.69^{+0.44}_{-0.26}$ & $1.56^{+0.21}_{-0.16}$ & $9.31^{+1.35}_{-0.79}$ & $1.71^{+0.50}_{-0.29}$ & $1.69^{+0.42}_{-0.27}$ & $9.33^{+1.23}_{-0.80}$\\
$W$/d.o.f. & 68.3 / 81 & 99.8 / 95 & 74.7 / 81 & 67.9 / 77 & 79.3 / 91 & 74.5 / 77\\
\hline
\multicolumn{1}{c}{Spectral model} & \multicolumn{6}{c}{\tablenotemark{\small a}\tt tbabs*tbpcf*(bbodyrad+gauss)}\\
$N_\mathrm{H,pc}$ ($10^{22}$ cm$^{-2}$)\tablenotemark{\small b} & $10.4^{+2.3}_{-2.0}$ & $6.8^{+0.9}_{-0.9}$ & $5.4^{+0.9}_{-0.8}$ & $10.5^{+2.3}_{-1.9}$ & $7.8^{+1.0}_{-0.9}$ & $5.4^{+0.9}_{-0.8}$\\
$1-f_\mathrm{pc}$ ($10^{-2}$) & $1.79^{+1.41}_{-1.17}$ & $6.44^{+1.78}_{-1.51}$ & $<1.17$ & $<1.59$ & $<0.87$ & $<1.21$\\
$T_\mathrm{BB}$ (keV) & $2.27^{+0.38}_{-0.29}$ & $2.05^{+0.17}_{-0.15}$ & $2.38^{+0.36}_{-0.28}$ & $2.18^{+0.37}_{-0.28}$ & $1.61^{+0.14}_{-0.12}$ & $2.39^{+0.36}_{-0.28}$\\
$R_\mathrm{BB}$ (km) & $0.71^{+0.18}_{-0.15}$ & $0.76^{+0.10}_{-0.09}$ & $1.67^{+0.33}_{-0.28}$ & $0.75^{+0.20}_{-0.16}$ & $1.05^{+0.18}_{-0.15}$ & $1.65^{+0.33}_{-0.28}$\\
$F_\mathrm{FeK\alpha}$ ($10^{-11}$ {\uflux})\tablenotemark{\small c} & $<2.0$ & $1.7^{+1.0}_{-0.9}$ & $<10.7$ & $<2.1$ & $2.3^{+1.0}_{-0.9}$ & $<10.6$\\
$EW_\mathrm{FeK\alpha}$ (eV)\tablenotemark{\small c} & $<145$ & $138^{+78}_{-75}$ & $<123$ & $<152$ & $228^{+132}_{-83}$ & $<116$\\
$F_\mathrm{abs}$  ($10^{-9}$ {\uflux})\tablenotemark{\small d} & $0.69^{+0.06}_{-0.06}$ & $0.65^{+0.04}_{-0.03}$ & $5.06^{+0.47}_{-0.43}$ & $0.65^{+0.06}_{-0.06}$ & $0.49^{+0.03}_{-0.03}$ & $5.09^{+0.46}_{-0.42}$\\
$F_\mathrm{unabs}$ ($10^{-9}$ {\uflux})\tablenotemark{\small d} & $1.05^{+0.10}_{-0.09}$ & $0.89^{+0.05}_{-0.04}$ & $6.58^{+0.56}_{-0.48}$ & $1.02^{+0.11}_{-0.05}$ & $0.78^{+0.05}_{-0.04}$ & $6.60^{+0.55}_{-0.48}$\\
$L_\mathrm{unabs}$ ($10^{36}$ {\ulumi})\tablenotemark{\small e} & $1.16^{+0.11}_{-0.10}$ & $0.98^{+0.05}_{-0.05}$ & $7.23^{+0.62}_{-0.52}$ & $1.12^{+0.12}_{-0.06}$ & $0.85^{+0.06}_{-0.04}$ & $7.25^{+0.61}_{-0.52}$\\
$W$/d.o.f. & 68.6 / 81 & 144.0 / 95 & 78.3 / 81 & 67.3 / 77 & 81.9 / 91 & 78.3 / 77\\
\hline
\enddata
\tablenotetext{\small a}{$N_\mathrm{H}$ of {\tt tbabs} model for ISM is fixed at the Galactic H$_\mathrm{I}$ density ($=0.55\times 10^{22}$ {\ucol}).}
\tablenotetext{\small b}{$N_\mathrm{H}$ of {\tt tbpcf} model for partial covering absorber} 
\tablenotetext{\small c}{Flux ($F_\mathrm{FeK\alpha}$) and equivalent width ($EW_\mathrm{FeK\alpha}$) of iron-K$\alpha$ 6.4 keV line modeled with a narrow ($\sigma=0.1$ keV) Gaussian}
\tablenotetext{\small d}{Absorbed ($F_\mathrm{abs}$) or unabsorbed ($F_\mathrm{unabs}$) 0.5-10 keV flux}
\tablenotetext{\small e}{Unabsorbed 0.5-10 keV lminosity assuming isotropic emission located at 3 kpc distance} 
\end{deluxetable*}

\section{Discussions} \label{sec:discussion}
MAXI J0709 is a new X-ray transient discovered on 2022 January 25.
The transient outburst behavior characterized by
the short duration of a few hours,
rapid variability with a time scale of $\lesssim 1$ s,
and large luminosity swing from
$10^{32}$ {\ulumi} to $10^{37}$ {\ulumi} 
agree with that of the typical SFXTs
\citep{sugizaki2022}.
The heavy X-ray absorption of $N_\mathrm{H}\sim10^{22}\mathord{-}10^{23}$ {\ucol}
changing during the short outburst period
is not typical 
but similar spectral changes has been reported for a few SFXTs
\citep[e.g.][]{2011A&A...531A.130B}.
On the other hand,
the optical companion, LY CMa,
exhibits properties of both a Be star as well as a B supergiant.
Therefore, the mechanism of mass accretion onto the compact object,
which is presumably a magnetized NS, 
has been under debate \citep{sugizaki2022}.
The results of the NICER data analysis in the previous section
reveal the extreme variability feature 
in more detail.
We here consider  possible mechanisms that can explain
all of the observed results.

\subsection{Short-duration rapid-flaring activity} 
\label{sec:discuss_lc}
The NICER light curves in figures \ref{fig:lcmxnr} and
\ref{fig:lcmulti} reveal that the X-ray flare activity
continued to appear until $t\sim 7$ h from the first MAXI/GSC detection at $t=0$,
but its duty cycle 
is limited to a few short segments 
including the three remarkable flare periods of A: $\sim$50 s, C1: $\sim$50 s, and C2: $\sim$100 s 
(see figure \ref{fig:lcmulti}).
Since the total NICER exposure from the observation start at $t=3$ h to 7 h  is  $3269$ s,
the flare duty cycle 
during this 4 hours
is $\sim (50+50+100 \,\mathrm{s} )/ 3269 \,\mathrm{s}\simeq 0.06$.
The duty cycle of the initial 3 hours from  $t=0$ to 3 h covered by the MAXI/GSC survey
is roughly estimated to be $\sim 0.5$ 
from the significant X-ray detections at the first ($t=0$) and third ($t\simeq$ 3 h) scans
and the non-detection at the second scan ($t\simeq 1.5$ h).
Therefore, the duty cycle of the flare activity during the entire 7-h outburst period,
$\delta_\mathrm{fl}$, 
can be roughly estimated  as 
%
\begin{equation}
\delta_\mathrm{fl} \sim \frac{0.5\times 3 +0.06\times 4}{7}\simeq 0.25.
\end{equation}

The overall outburst profile consisting of
very sparse, short-lived ($\lesssim 100$ s), flaring periods
seems even extreme compared to those of other typical SFXT outbursts 
\citep[e.g.][]{2013AdSpR..52.1593R,2017A&A...608A.128B}.
The fine light-curve profile in figure \ref{fig:lcmulti}  
shows that each flare consists of short spikes, 
each lasting a few seconds and reaching typically $\sim100$ cts s$^{-1}$ at the peak. 
Assuming the spectral model of a cutoff power-law continuum of $\Gamma\simeq1$ and $E_\mathrm{fold}=22$ keV
obtained in section \ref{sec:ana_spec},
the count rate $\sim100$ cts s$^{-1}$ corresponds to 
the absorption-corrected bolometric luminosity of $\sim 1.8\times 10^{37}$ {\ulumi}.
Therefore,
the maximum count rate of 1000 cts s$^{-1}$ in the data of every 0.1-s time bin  
indicates that the source luminosity reached up to $\sim 1.8\times 10^{38}$ {\ulumi}
during the peak 0.1 seconds. 
It is almost equal to the Eddington limit for a typical $1.4 M_\sun$ NS
and comparable to those of the brightest SFXTs
that have ever been reported
\citep[e.g.][]{2015AdSpR..55.1255B,2015A&A...576L...4R}.

\subsection{Local absorption medium and iron-K$\alpha$ line} 

In the X-ray spectral analysis in section \ref{sec:ana_spec},
we found that
each flare spectrum is represented by a power-law of $\Gamma\simeq 1$
with a heavy partial-covering absorption of
$N_\mathrm{H,pc}= (8\mathord\sim14)\times 10^{22}$ {\ucol}
and $1-f_\mathrm{pc}\lesssim 0.02$.
The 6.4 keV iron-K$\alpha$ line, 
suggesting the emission from neutral iron,
with $EW_\mathrm{FeK\alpha}\simeq 100$ eV
was detected in the C2* spectrum, which is collected from the flare C2 except for the brightest 4 s (designated C2a).
These power-law model and iron-K$\alpha$ line parameters agree well with those of other HMXBs including SFXTs
\citep[e.g.][]{2018A&A...610A..50P}.
The $EW_\mathrm{FeK\alpha}$ to $N_\mathrm{H,pc}$ ratio
$\sim 100 $ eV / $10^{23}$ {\ucol}
suggests that the CSM distribution around the primary X-ray source
is not completely uniform,
in which case the ratio should be $\simeq 30$ eV / $10^{23}$ {\ucol}
\citep[][]{1985SSRv...40..317I,2010ApJ...715..947T}.
The discrepancy between $\Gamma\sim 1$ obtained here
and $\Gamma\sim2$ obtained from the MAXI spectrum for a possible continuous flare
\citep{sugizaki2022}
suggests a cutoff or break in 
the simple power-law model at $\sim 10$ keV.
We fitted the MAXI spectrum with an exponential cutoff power-law model
and obtained the best-fit $E_\mathrm{fold}=22^{+\infty}_{-13}$ keV,
which is consistent with that in typical HMXBs
\citep[e.g.][]{1999ApJ...525..978M,2002ApJ...580..394C}.

\begin{figure}
\begin{center}
\includegraphics[width=0.48\textwidth]{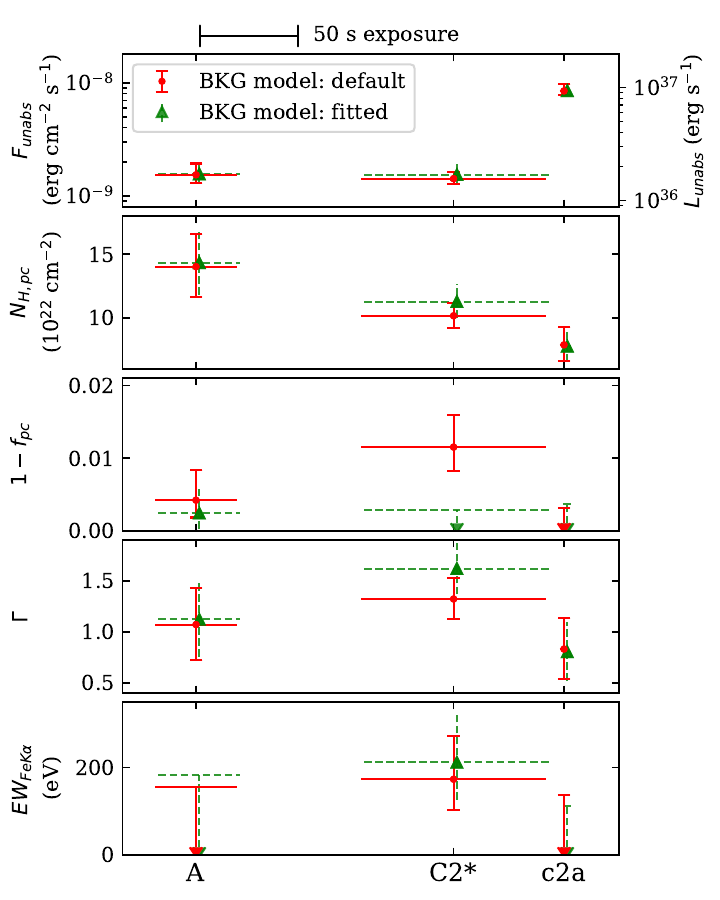}
\caption{
  Best-fit spectral parameters,
  $F_\mathrm{unabs}$ ($L_\mathrm{unabs}$), $\Gamma$, $N_\mathrm{H,pc}$, $1-f_\mathrm{pc}$,
  and $EW_\mathrm{FeK\alpha}$ 
  of a power-law continuum model for three flaring periods, A, C2*, and C2a, listed in table \ref{tab:specparams},
  assuming background models of SCORPEON default (red) and that fitted (green)
  shown in figure \ref{fig:espec}.
  The vertical error bars represent the 90\% confidence limits of the statistical uncertainties.
  The length of horizontal bars represents the exposure time of each data in a scale given at the top.
}
\label{fig:specparam}
\end{center}
\end{figure}

Figure \ref{fig:specparam}
shows the changes in the obtained best-fit parameters of the partial-covering-absorption power-law model,
$F_\mathrm{unabs}$, $N_\mathrm{H,pc}$, $1-f_\mathrm{pc}$, $\Gamma$, and $EW_\mathrm{FeK\alpha}$
among the three flare periods of A, C2*, and C2a 
in table \ref{tab:specparams}.
$N_\mathrm{H,pc}$ decreased from the flare A to C2* and C2a,
consistent with the hardness-ratio decrease from A to C2* and C2a in figure \ref{fig:hid}.
During the flare C2,
$F_\mathrm{unabs}$ increased from C2* to C2a,
while all other parameters, $N_\mathrm{H,pc}$, $1-f_\mathrm{pc}$, $\Gamma$, and $EW_\mathrm{FeK\alpha}$,
apparently decreased.
If the SCORPEON background model parameters are allowed to vary
from the default model values,
the change in $1-f_\mathrm{pc}$ becomes insignificant
but the changes in $N_\mathrm{H,pc}$, $\Gamma$, and $EW_\mathrm{FeK\alpha}$ become more significant.
We estimated the significance of these $N_\mathrm{H,pc}$, $\Gamma$, and $EW_\mathrm{FeK\alpha}$ changes
to be greater than 95\%, 95\%, and 90\%, respectively, by $\chi^2$ tests,
for both background models of the default SCORPEON parameters and those fitted within the allowed parameter ranges.

\begin{figure*}
  \begin{center}
  \includegraphics[width=0.8\textwidth]{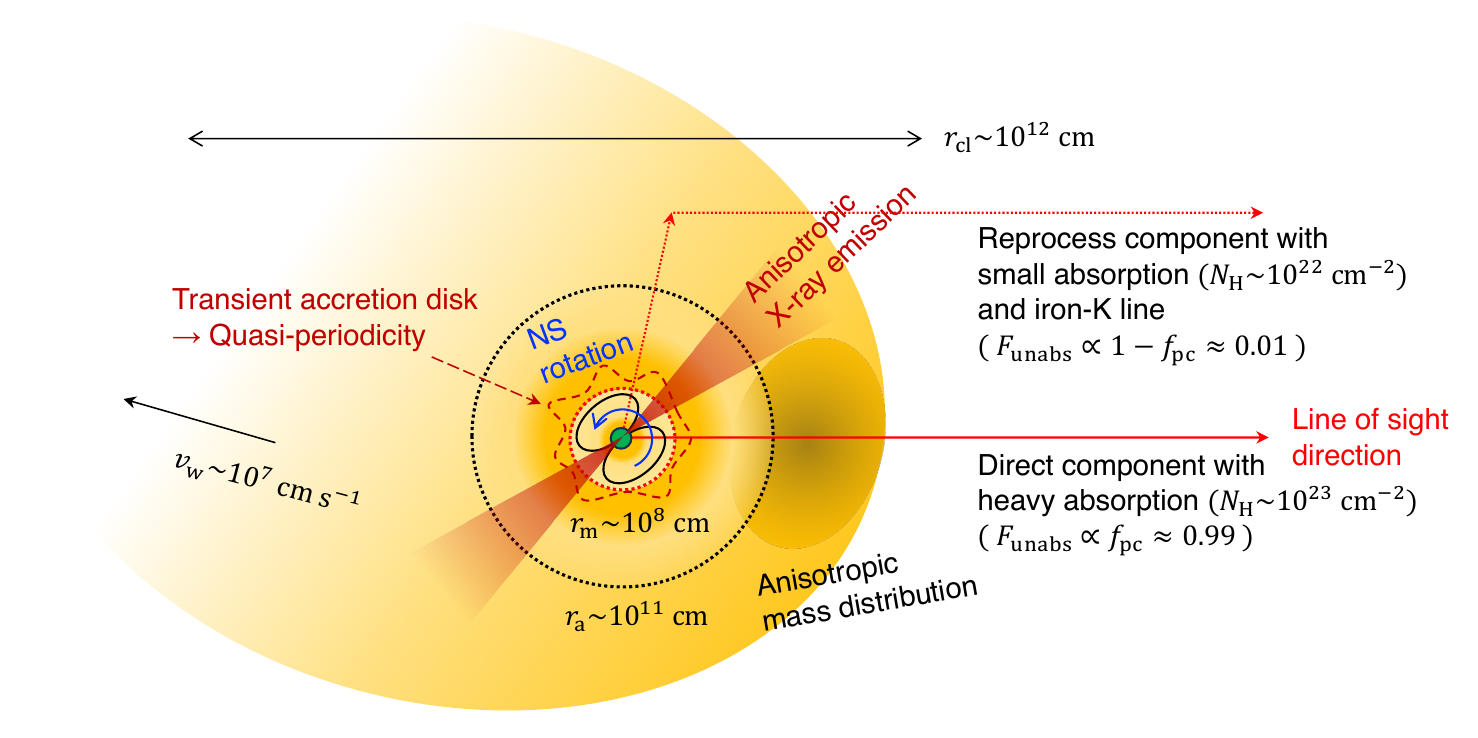}
  \caption{Schematic picture of compact X-ray object and surrounding CSM distribution
    suspected from X-ray spectral changes and rapid flaring activities
    during the short outburst based on the clumpy wind accretion scenario.}
  \label{fig:sgxb}
\end{center}
\end{figure*}

The $N_\mathrm{H,pc}$ and $EW_\mathrm{FeK\alpha}$ 
give us information on the CSM distribution around the primary X-ray source.
The anti-correlations 
between $F_\mathrm{unabs}$ and $EW_\mathrm{FeK\alpha}$ 
can be naturally explained by a scenario that the primary X-ray emission
from the compact X-ray object is anisotropic
and the beamed emission direction was aligned to the line of sight 
at the time of the flare peak (C2a),
where the relative intensity of the direct component dominating 
the continuum spectrum increased but 
that of the reprocess component by the surrounding CSM, 
which is a major source of iron-K$\alpha$ emission line,
decreased.
In figure \ref{fig:sgxb}, the overall schematic picture is illustrated.
Although, alternative scenarios cannot be ruled out.

\subsection{Application to clumpy wind accretion model}
\label{sec:clacc}

The outburst light curve discussed above, 
characterized by short, sparse, rapid-flaring activity,
is different from that of typical BeXBs \citep[e.g.][]{2011Ap&SS.332....1R}
and leads us to the idea of mass accretion via clumpy stellar winds 
that has been considered for SgXBs
\citep[e.g.][]{2016A&A...589A.102B}.
In fact,
the outburst duration of $\sim 7$ h ($= 2\times 10^{4}$ s)
and spectral variation due to $N_\mathrm{H}$ change from $10^{22}$ to $10^{23}$ {\ucol}
are quite similar to those observed previously in a typical SFXT, IGR J18410$-$0535 \citep{2011A&A...531A.130B},
although the peak luminosity ($\gtrsim 10^{37}$ {\ulumi}) is significantly higher than 
in the  IGR J18410$-$0535 case ($\sim 10^{35}$ {\ulumi}).
Hence, following the previous SFXT example.
we examine the present results with a wind-accretion scenario
that has not been considered for BeXBs.
We denote the mass and radius of the compact X-ray object, which is likely to be a NS, by
$M_\mathrm{NS}$ and $R_\mathrm{NS}$, and those of the mass-donating stellar companion LY CMa 
by $M_{*}$ and $R_{*}$, 
hereafter.

The radial extent of the clump $r_\mathrm{cl}$ is 
roughly 
estimated from 
the outburst duration $t_\mathrm{ob}$ 
and relative velocity $v_\mathrm{rel}$ between the clump  
and the NS
as 
\begin{equation}
  r_\mathrm{cl} \simeq v_\mathrm{rel} t_\mathrm{ob}.
  = 10^{12} v_\mathrm{rel8} t_\mathrm{ob4} \;\mathrm{cm}
  \label{equ:rcl}
\end{equation}
where $t_\mathrm{ob4} = t_\mathrm{ob} /10^{4}\,\mathrm{s}$
and $v_\mathrm{rel8} = v_\mathrm{rel} /10^{8}\,\mathrm{cm\,s^{-1}}$.
Because the clump moves with the stellar wind,
its radial velocity from the donor star LY CMa is approximated by the wind velocity $v_\mathrm{w}$.
\citet{bhattacharyya2022} derived the effective temperature $T_\mathrm{eff}=20$ kK and 
$R_{*}\simeq 11.8 R_\sun = 8.2\times 10^{11}$ cm
from the optical SED model fit assuming the object to be
an evolved Be star with a surface gravity $\log g = 3$.
These $R_{*}$ and $\log g$ indicate $M_{*} = 5M_\sun$, which agrees with the typical Be-star mass, 
and the surface escape velocity of $v_\mathrm{esc}=4.0\times 10^{7}$ cm s$^{-1}$. 
%
The terminal wind velocity $v_\infty$ is then estimated to be  $\sim (4{-}8)\times 10^{7}$ cm s$^{-1}$
from $v_\infty/v_\mathrm{esc}\sim1{-}2$  in B1.5  Be stars \citep[][]{1989MNRAS.241..721P} and B supergiants \citep[][]{2006A&A...446..279C},
which is slower than $v_\infty\gtrsim 10^{8}$ cm s$^{-1}$ in typical SgXBs \citep{2020A&A...634A..49H}.
Meanwhile,
the orbital velocity $v_\mathrm{orb}$  of the compact X-ray object around the primary star LY CMa
and the distance between the two objects $r_\mathrm{orb}$ 
are expressed using the orbital parameters, orbital period $P_\mathrm{orb}$ and eccentricity $e$, as
\begin{eqnarray}
r_\mathrm{orb} &=& \left\{\frac{G\left(M_\mathrm{NS}+M_{*}\right)P_\mathrm{orb}^2}{4\pi^2} \right\}^{1/3} (1\pm e)\nonumber\\
&=& 2.5\times 10^{12} P_\mathrm{10d}^{2/3} M_{6.4}^{1/3} (1\pm e)\; \mathrm{cm} , \label{equ:rorb}\\
v_\mathrm{orb} &=& \left\{ \frac{2\pi G\left(M_\mathrm{NS}+M_{*}\right)} {P_\mathrm{orb}} \right\}^{1/3}  \frac{\sqrt{1-e^2}}{1\pm e} \nonumber\\
&=& 1.8\times 10^7 P_\mathrm{10d}^{-1/3} M_{6.4}^{1/3} \frac{\sqrt{1-e^2}}{1\pm e}  \; \mathrm{cm\ s^{-1}} \label{equ:vorb} 
\end{eqnarray} 
where $P_\mathrm{10d} = P_\mathrm{orb}/$ 10 d, $M_{6.4} =  (M_\mathrm{NS}+M_{*})/ (1.4M_\sun + 5 M_\sun)$,
and the $\pm$ signs represent equations for periastron ($-$) and apastron ($+$) epochs, respectively.
Then, assuming the standard wind-velocity law
so called $\beta$-velocity law 
\begin{equation}
v_\mathrm{w} = v_\infty \left(1-\frac{R_{*}}{r_\mathrm{orb}} \right)^\beta 
\label{equ:betalaw}
\end{equation} 
%
%
where $\beta\sim 0.5$ in a classical line-driven wind model \citep{1975ApJ...195..157C}
has been revised to $\beta\sim 0.8{-}2$ with hydrodynamical calculations and observations \citep[e.g.][]{2006A&A...446..279C,2011A&A...534A..97K},
we can calculate $v_\mathrm{w}$ from $v_\infty$, $R_{*}$ and $r_\mathrm{orb}$ estimated above
and  the orbital parameters, $P_\mathrm{orb}$ and $e$, which are unknown.
Equation (\ref{equ:betalaw}) means that
the larger $r_\mathrm{orb}$ is, 
the closer $v_\mathrm{w}$ is to  $v_\infty$.
Assuming $P_\mathrm{orb}\simeq10$ d and $e\lesssim 0.4$ (not extremely eccentric)
as typical in BeXBs \citep[e.g.][]{2019NewAR..8601546K},
we can roughly estimate from equations (\ref{equ:rorb})(\ref{equ:vorb})(\ref{equ:betalaw})
that $r_\mathrm{orb}\sim 2.5\times 10^{12}$ cm,
$v_\mathrm{orb}\sim 1.8\times 10^7$ cm s$^{-1}$,  
and thus $v_\mathrm{w}\sim 0.5^\beta v_\infty \sim (1{-}4)\times 10^7$ cm s$^{-1}$.
Therefore, we expect mostly $v_\mathrm{orb} \lesssim v_\mathrm{w}$
in typical binary conditions as discussed.
Given these expected ranges of $v_\mathrm{orb}$ and $v_\mathrm{w}$,
we can estimate $v_\mathrm{rel}\simeq\sqrt{v_\mathrm{orb}^2 + v_\mathrm{w}^2}$ to be $\sim (2.5{-}6) \times 10^{7}$ cm s$^{-1}$.

When the NS enters the clump,
the matter within the accretion radius $r_\mathrm{a}$, 
\begin{eqnarray}
  r_\mathrm{a}
  &=& 2GM_\mathrm{NS} / v_\mathrm{rel}^2 \nonumber \\
  &=& 3.7\times 10^{10} M_\mathrm{1.4} v_\mathrm{rel8}^{-2}  \; \mathrm{cm}
  \label{equ:racc}
\end{eqnarray}
is accreted \citep{1952MNRAS.112..195B}.
Assuming that the mass distribution in the clump is approximately uniform,
the relation between the total clump mass 
$M_\mathrm{cl}$ and the accretion mass $M_\mathrm{acc}$
is given by
\begin{equation}
M_\mathrm{cl} = (r_\mathrm{cl}/r_\mathrm{a})^2 M_\mathrm{acc}.
\label{equ:mcl}
\end{equation}
Meanwhile, $M_\mathrm{acc}$ can be estimated from the outburst light curve
representing the luminosity $L(t)$ as a function of time $t$
by assuming that the X-ray emission is powered by mass accretion onto the NS surface,
i.e.
\begin{equation}
 GM_\mathrm{NS} M_\mathrm{acc} / R_\mathrm{NS} =  \int  L(t) dt = L_\mathrm{fl} \delta_\mathrm{fl} t_\mathrm{ob}\nonumber\\
\label{equ:macc}
\end{equation}
where  $L_\mathrm{fl}$ is the average luminosity 
during the flaring duty cycle ($=\delta_\mathrm{fl} t_\mathrm{ob}\nonumber$)
out of the entire outburst period, i.e. $L_\mathrm{fl}\sim 10^{36}$ {\ulumi}. 
From equations (\ref{equ:rcl})(\ref{equ:racc})(\ref{equ:mcl})(\ref{equ:macc}), we obtain 
\begin{eqnarray}
  M_\mathrm{cl} &=& \frac{R_\mathrm{NS} L_\mathrm{fl} \delta_\mathrm{fl} t_\mathrm{ob}^3 v_\mathrm{rel}^6}{\left(GM_\mathrm{NS}\right)^3}\nonumber\\
  &\simeq& 1.6\times 10^{22} M_\mathrm{1.4}^{-3} R_\mathrm{6}  L_\mathrm{36} \delta_{0.1} t_\mathrm{ob4}^{3} v_\mathrm{rel8}^{6} \;\mathrm{g},
  \label{equ:mcl2}
\end{eqnarray}
where $L_\mathrm{36} = L_\mathrm{fl}/10^{36}$ {\ulumi} and $\delta_{0.1} = \delta_\mathrm{fl}/0.1$.
Then,
the hydrogen column density through the clump $N_\mathrm{H,cl}$ is represented by 
\begin{eqnarray}
  N_\mathrm{H,cl} &=& \frac{M_\mathrm{cl}}{r_\mathrm{cl}^2 m_\mathrm{p}} =  \frac{R_\mathrm{NS} L_\mathrm{fl} \delta_\mathrm{fl} t_\mathrm{fl} v_\mathrm{rel}^4}{\left(GM_\mathrm{NS}\right)^3}\nonumber\\
  &\simeq& 0.9\times 10^{22} M_{1.4}^{-3} R_{6} L_{36} \delta_{0.1} t_\mathrm{ob4} v_\mathrm{rel8}^{4} \;\mathrm{cm^{-2}},
  \label{equ:nhcl}
\end{eqnarray}
where $m_\mathrm{p}$ is the proton mass.
Applying the observed parameters, $L_{36}\sim 5$, $\delta_{0.1}\sim 2.5$, and $t_\mathrm{ob4}\sim 2$
obtained from the present data analysis,
and $v_\mathrm{rel8}\sim 0.25{-}0.6$ estimated above 
into the equations (\ref{equ:mcl2}) and (\ref{equ:nhcl}),
we obtain $M_\mathrm{cl} \sim (0.04{-}7)\times 10^{22}$ g and $N_\mathrm{H,cl} \sim (0.09{-}3)\times 10^{22}$ {\ucol}.
The calculated $N_\mathrm{H,cl}$ 
is smaller than the absorption column density  $N_\mathrm{H,pc}\sim 1\times 10^{23}$  {\ucol} 
for the X-ray source determined by the spectral analysis (section \ref{sec:ana_spec}).
On the other hand, the $M_\mathrm{cl}\sim 10^{22}$ g  is on the larger side among those of typical SgXBs, 
which are $\sim 10^{18}$ to $10^{22}$ g \citep[e.g.][]{2017SSRv..212...59M}.
These discrepancies suggest that the clump mass distribution is not spherically uniform as assumed in equation (\ref{equ:mcl}),
but may be like a filament or a cluster of multiple clumps.

In figure \ref{fig:sgxb}, the scales of $r_\mathrm{cl}\sim10^{12}$ cm
and $r_\mathrm{a}\sim 10^{11}$ cm are schematically presented.

\subsection{Quasi-periodic variation and transient accretion-disk scenario}
\label{sec:disuss_qpo}

The PDS in figure \ref{fig:pds} reveals characteristic structures 
below $\sim0.5$ Hz during the flare activities
and also an isolated peak at 1.1 Hz during the brightest flare of $L\simeq 10^{37}$ {\ulumi}.
Because the PDS peak has a width typical of QPOs,
it is unlikely to be a coherent pulsation due to the NS spin.
What does the QPO represent?

QPOs have been reported in a number of HMXB pulsars, 
mostly BeXB pulsars
\citep[e.g.][]{1989ApJ...346..906A,1994ApJ...436..871T,1996ApJ...459..288F,2010MNRAS.407..285J,2011MNRAS.417..348D,2019ApJ...872...33R,2023ApJ...954...48P}
but also a few SgXB pulsars
\citep[e.g.][]{2022MNRAS.512.4792R,2022MNRAS.516.5579L}.
%
%
%
%
In these HMXB pulsars, QPOs appeared at frequencies from
1 mHz to 1.2 Hz \citep{2010MNRAS.407..285J}.
The present QPO peak of 1.1 Hz
is close to the highest one of 1.2 Hz observed in XTE J0111.2$-$7317 \citep{2007ApJ...660.1409K}.
The periodicities of the QPOs in the HMXB pulsars have been considered mainly
from the Keplerian rotation frequency at the inner edge of the accretion disk
interacting with the NS magnetosphere
\citep[Keplerran Frequency Model; KFM;][]{1987ApJ...316..411V}
or its beat frequency with the NS spin frequency
\citep[Beat Frequency Model; BFM;][]{1985Natur.316..239A}.
These models successfully explain some of the QPOs observed in BeXBs \citep[e.g.][]{1996ApJ...459..288F},
but the attempt to explain all of these features with a unified picture has not been successful
\citep{2009A&A...493..809B}.
The discrepancy between the observed QPO frequencies
and those predicted by KFM or BFM assuming the NS magnetic fields
measured with the cyclotron resonance features
suggests that
the QPO model predictions should have an error of about one order of magnitude
\citep[e.g.][]{2010MNRAS.407..285J}.

Possible QPOs have been also reported in a few SFXTs.
IGR J17544$-$2619 
showed a broad PDS peak at 86 mHz (period $=11.6$ s) 
with the quality factor $Q\simeq 3$ like QPO during the bright flare  
of the peak luminosity $\sim3\times 10^{38}$ {\ulumi}
on 2014 October 10 \citep{2015A&A...576L...4R}, but
its periodicity has never been confirmed in the later observations
\citep{2016A&A...596A..16B}.
%
%
%
%
Also, IGR J19140+0951, a possible intermediate HMXB classified between classical persistent SgXBs and SFXTs,
has shown transient QPOs at 1.46 mHz and 0.17 mHz, 
the latter of which is considered to represent the NS spin \citep{2016MNRAS.460.3637S}.
These irreproducible observation results suggest that
their origins would be irregular transient events, such as
transient accretion-disk formation  
\citep[][]{2015A&A...576L...4R,2016MNRAS.460.3637S}.

We here consider the transient accretion-disk scenario 
in which QPOs are expected by the KFM  or BFM mechanism,
following the previous examples mentioned above.
As discussed in section \ref{sec:clacc},  
the wind velocity of LY CMa is expected to be slower than that of typical SgXBs,
suggesting a higher mass-capture rate ($r_\mathrm{a}^2v_\mathrm{rel} \propto  v_\mathrm{rel}^{-3}$)
that would favor the accretion-disk formation \citep[e.g.][]{2010MNRAS.408.1540D}. 
Assuming that the captured accreting matter conserves its angular momentum as it falls onto the magnetized NS, 
sustained accretion-disk formation outside the NS magnetosphere requires  $v_\mathrm{rel}\lesssim 3.5\times 10^{7}$ cm s$^{-1}$ \citep{1981A&A...102...36W}
under a typical binary orbit introduced at equation (\ref{equ:rorb}),
which corresponds to the lower end of the estimated possible range $v_\mathrm{ref}\simeq(2.5{-}6)\times 10^{7}$ cm s$^{-1}$
in section \ref{sec:clacc}.
On the other hand, transient disk formation has been proposed  
for flare-like events observed in the typical SgXB Vela X-1 with $v_\mathrm{ref}\gtrsim 10^{8}$ cm s$^{-1}$,
where the inhomogeneous wind density and velocity induce the instability 
leading to the disk formation and dissipation
\citep{2008A&A...492..511K}.
In fact,
the observed outburst duration $t_\mathrm{ob}\sim 10^{4}$ s is on the same order of magnitude as the model prediction $\sim 6GM_\mathrm{NS} v_\mathrm{rel}^{-3}$ 
\citep{1988ApJ...331L.117T}.
Therefore, the transient disk formation scenario is considered to be possible for any $v_\mathrm{ref}$ within the expected range. 

The radius $r_\mathrm{K}$ around the NS
with a Keplerian orbital frequency $\nu_\mathrm{K}$ is given by
\begin{eqnarray}
r_\mathrm{K} &=& \left(\frac{GM_\mathrm{NS}}{4\pi^2v_\mathrm{K}^2}\right)^{1/3} \nonumber\\
&=& 1.7\times 10^{8} M_\mathrm{1.4}^{1/3} \nu_\mathrm{K}^{-2/3} \; \mathrm{cm}.  
  \label{equ:rk}
\end{eqnarray}
In KFM, 
the QPO frequency $\nu_\mathrm{qpo} =1.1$ Hz 
is expected to be equal to $\nu_\mathrm{K}$ at the inner-disk radius,
where the NS spin frequency $\nu_\mathrm{s}$ should be  $<\nu_\mathrm{K} =1.1$ Hz
because the mass accretion was not inhibited by centrifugal barrier (propeller effect).
In BFM, 
$\nu_\mathrm{qpo}$ represents the beat frequency between the $\nu_\mathrm{K}$ 
and $\nu_\mathrm{s}$,
i.e. $\nu_\mathrm{qpo} = \nu_\mathrm{K} - \nu_\mathrm{s}$.  
The $\nu_\mathrm{s}$ has not been known in the present data.
The PDS in figure \ref{fig:pds} shows
no significant coherent pulsation 
in the 2.5 mHz -- 1 kHz range. 
Considering that 
most of the HMXB pulsars that have been observed 
have spin periods  $P_\mathrm{s} \gtrsim1$ s,
\citep[e.g.][]{2019NewAR..8601546K},
we can expect likely $\nu_\mathrm{s}\lesssim 1$ Hz,
and thus $\nu_\mathrm{K} = \nu_\mathrm{qpo}  +\nu_\mathrm{s} \sim$ 1--2  Hz.
Hence, we here focus on the case of $\nu_\mathrm{K}\sim 1$ Hz
at the inner-disk radius.
We consider the exception case of of $\nu_\mathrm{K}\gg 1$ Hz in the next section.
From equation (\ref{equ:rk}),
the inner-disk radius $r_\mathrm{K}$ for $\nu_\mathrm{K}\sim 1$ Hz 
is estimated to be $\sim 10^{8}$ cm,
which is much smaller than $r_\mathrm{a}\sim 10^{11}$ cm in equation (\ref{equ:racc}),
suggesting that the accreting matter flow would become
roughly spherically uniform within $r_\mathrm{a}$.
The schematic picture is shown on figure \ref{fig:sgxb}.

The magnetospheric radius $r_\mathrm{m}$ at which the pressure of the NS magnetic field
balances the ram pressure of the accreting matter 
is expressed by 
\begin{equation}
  r_\mathrm{m} = \zeta \left(\frac{\mu^4}{2GM_\mathrm{NS}\dot{M}_\mathrm{acc}^2}\right)^{1/7} \label{equ:ralfven}
\end{equation}
where $\mu$ 
is the NS magnetic dipole moment and 
$\zeta (\sim1)$ is the geometrical factor of the accreting matter flow;
$\zeta=1$ in spherical accretion
and $\zeta=0.52$ in the disk accretion model proposed by \citet{GhoshLamb1979II,GhoshLamb1979III}.
Assuming that the accreting matter releases all the energy on the NS
surface with X-ray radiation, i.e.
$L=GM_\mathrm{NS}\dot{M}_\mathrm{acc}/R_\mathrm{NS}$,
equation (\ref{equ:ralfven}) is reduced to
\begin{equation}
  r_\mathrm{m} = 2.7\times 10^{8} \zeta\mu_{30}^{4/7} M_{1.4}^{1/7} R_{6}^{-2/7} L_{37}^{-2/7} \; \mathrm{cm},
  \label{equ:rm}  
\end{equation}
where $\mu_{30} = \mu/10^{30}$ G cm$^{-3}$ and $L_{37} = L / 10^{37}$ {\ulumi}.
If $r_\mathrm{K}=r_\mathrm{m}$ as supposed in both KFM and BFM,
we obtain the equation of $\mu_{30}$ from equations (\ref{equ:rk}) and (\ref{equ:rm}),
\begin{equation}
  \mu_{30} =  0.44   \zeta^{-7/4} M_{1.4}^{1/3} R_{6}^{1/2} L_{37}^{1/2} \nu_\mathrm{K}^{-7/6} .
  \label{equ:mu30}
\end{equation}
Hence, the NS surface magnetic field $B_\mathrm{s}= 10^{12} \mu_{30} R_\mathrm{6}^3$ G 
can be estimated from equation (\ref{equ:mu30}).
Applying the observed parameters
of $L\sim 10^{37}$ {\ulumi} and $\nu_\mathrm{K}\simeq 1$ Hz,
we obtain
$B_\mathrm{s}\sim0.4\times 10^{12}$ G in the spherical accretion case ($\zeta=1$)
and $\sim2\times 10^{12}$ G in the disk accretion case ($\zeta\simeq0.5$).
This agrees with the typical $B_\mathrm{s}$ of HMXB pulsars
$(1\mathord{\sim}10) \times 10^{12}$ G
measured by cyclotron resonance features
\citep[e.g.][]{1999ApJ...525..978M}.

\subsection{What makes MAXI J0709 different from other HXMB systems?} 
As discussed above,
the observed short outburst duration and fast variability 
are quite unique  compared to those of other well-known HMXBs.
The X-ray time variation
is considered to reflect the manner of mass accretion via CSM fed by the stellar companion. 
The optical spectroscopic observations suggest
the stellar companion LY CMa has a circumstellar Be disk whose structure changed
according to the X-ray activity
\citep[][]{bhattacharyya2022,2025arXiv250321118S}.
However, the X-ray light curve profile 
does not agree with  those of typical BeXBs, 
which usually 
cause outbursts lasting for several weeks to months
\citep[e.g.][]{2011Ap&SS.332....1R},
but resembles  those of SFXTs.
As discussed in section \ref{sec:clacc},
the overall profile is
largely explained by a clumpy wind accretion scenario, which has been often 
applied to  SFXTs. 
Therefore, we expect that the CSM condition around the compact X-ray object 
would be more similar to those of the SFXTs.
In fact, LY CMa is suggested to be an evolved Be star
located
between standard Be stars and B supergiants
on the color-magnitude diagram
\citep[][]{bhattacharyya2022}.
Similar objects that can be classified as intermediate SFXTs
have been reported \citep{2008A&A...492..163R,2014A&A...566A.131G}.

The very sparse flare activity ($\delta_\mathrm{fl}\sim 0.25$)
consisting of intermittent, short-lived ($\lesssim 100$ s), rapid-variability (times scale $\lesssim 1$ s) intervals during the outburst
(figures \ref{fig:lcmxnr}, \ref{fig:lcmulti})
seems extreme compared to typical SFXT outbursts. 
As estimated in section \ref{sec:clacc},
the wind velocity of LY CMa is thought to be slower than in typical SFXTs,
which may have some impact on the accretion process,
such as the formation of a transient accretion disk \citep[e.g.][]{2010MNRAS.408.1540D},
as discussed in section \ref{sec:disuss_qpo}.
Assuming that 
the compact object is a magnetized NS and 
the observed 1.1 Hz QPO during the brightest flare
is attributed to the inner boundary of the transient accretion disk
interacting with the NS magnetosphere at the radius $r_\mathrm{m}$,
the NS surface magnetic field is estimated 
to be $B_\mathrm{s}\sim 10^{12}$ G,
which agrees with those of typical HMXB pulsars
(section \ref{sec:disuss_qpo}).
The X-ray spectral results also  agree with those of typical HMXB pulsars,
whose X-ray emission is considered to originate from close to the magnetic poles on the NS surface
(section \ref{sec:ana_spec}).
%
No cyclotron-resonance features corresponding to the $B_\mathrm{s}$ 
have been detected in X-ray spectra obtained so far \citep{sugizaki2022}.

The NS spin frequency $\nu_\mathrm{s}$ has not been observed.
Based on the KFM scenario that 
$\nu_\mathrm{K}=\nu_\mathrm{qpo}=1.1$ Hz at $r_\mathrm{m}$,
we expect $\nu_\mathrm{s}< 1.1$ Hz, i.e. spin period $P_\mathrm{s}\gtrsim 1$ s
because the propeller effect did not inhibit mass accretion onto the NS surface.
In the BFM scenario,
it is possible that $\nu_\mathrm{s}>1$ Hz  
as long as $\nu_\mathrm{K}-\nu_\mathrm{s}=1.1$ Hz at $r_\mathrm{m}$.
If $\nu_\mathrm{s}\gg1$ Hz, i.e. a fast-rotating NS like a millisecond pulsar,
the NS magnetic field must be weak ($B_\mathrm{s} \ll 10^{12}$ G), 
as given by equation (\ref{equ:mu30}) for $\nu_\mathrm{K}>\nu_\mathrm{s}\gg1$ Hz at $r_\mathrm{m}$. 
This case is unlikely, because the observed X-ray light curve and spectral properties are different 
from those of X-ray binaries with weakly magnetized fast-rotating NS \citep[e.g.][]{2020arXiv201009005D}.
The absence of a coherent pulsation signal in the PDS from 2.5 mHz to 1 kHz (figure \ref{fig:pds})
may suggest that the NS spin axis is closer to the line of sight 
or that the $P_\mathrm{s}$ is longer than the individual observation segments ($\sim 1000$ s).

In the possible condition discussed above, where $\nu_\mathrm{s}< \nu_\mathrm{K}=1.1$ Hz at $r_\mathrm{m}$, 
the mass accretion is thought to be  in the subsonic propeller regime or subsonic settling regime 
where
the accreting matter is halted by the NS magnetosphere to form a transient accretion disk,
and then 
intermittently settling down onto the NS surface
\citep[e.g.][]{2008ApJ...683.1031B,2017SSRv..212...59M}.
The mechanism for the intermittent settling remains under debate
although some possible mechanisms have been proposed,
including Kelvin-Helmholtz instability \citep{2008ApJ...683.1031B}, 
radiative or Compton cooling \citep{2013MNRAS.428..670S},
and magnetic reconnection \citep{2014MNRAS.442.2325S}. 
The present data will help to resolve the issue.
Also, \citet{2008ApJ...683.1031B} proposed a hypothesis that 
SFXTs  would appear in HMXBs with 
strongly-magnetized NSs of
$B_\mathrm{s}\sim 10^{13}\mathord{-}10^{14}$ G like magnetars.
This disagrees with
the present estimate of $B_\mathrm{s}\sim 10^{12}$ G from the 1.1 Hz QPO assuming the KFM/BFM model,
but the hypothesis cannot be completely ruled out 
if the systematic uncertainty on the estimate using the QPO is considered.

\section{Conclusion}
We analyzed data of the NICER observations of the
new X-ray transient MAXI J0709$-$159 discovered on 2022 January 25.
The object has been identified as a HMXB with a stellar companion LY CMa,
which has a complex spectral type between Be stars and B supergiants.
The initial X-ray behavior observed by MAXI 
agrees well with that of typical SFXTs.

The NICER X-ray light curve from 3 hours to 7 days after the discovery
reveals that the X-ray activity continued until 7 hours 
in sparse short flares, each lasting $\lesssim 100$ s 
and reaching up to $\sim 1\times10^{38}$ {\ulumi}
at the instantaneous peak of 0.1 seconds.
The X-ray spectrum is represented by
a power-law continuum of $\Gamma\sim1$,
an iron-K$\alpha$ line at 6.4 keV
and a partial covering absorption of $N_\mathrm{H,pc}\sim 10^{23}$ {\ucol}
and $1-f_\mathrm{pc}\sim 0.01$, which are typical for HMXBs.
The spectral parameters were found to vary for each flare, and also during the brightest flare.
%
%
%
%
These light-curve and spectral feature
are in reasonable agreement with 
those expected from a clumpy wind-accretion scenario
as proposed for SgXBs.
In addition, the variability PDS shows
a broad peak at $\nu_0=1.1$ Hz with a width of $\Delta\simeq 0.11$ Hz like QPO.
If the QPO represents the Keplerian rotation 
at the inner edge of the accretion disk interacting with the NS magnetosphere
or its beat frequency with the NS spin,
the NS surface magnetic field $B_\mathrm{s}$ is estimated to be $\sim10^{12}$ G,
which is consistent with those of other typical HMXB pulsars.

What causes the extreme variability in  MAXI J0709
characterized by the sparse short flares
and the large peak luminosity during the outburst,
which is even extreme in the SFXT subclass,
is still unresolved.
One plausible scenario is
that the compact X-ray object is a (slow-rotating) 
magnetized NS of $P_\mathrm{s}\gtrsim 1$ s and
the mass accretion onto the NS
is in the subsonic propeller regime
or settling regime, where  accreting matter
is halted on the the NS magnetosphere and then
intermittently settling down onto the NS surface.
Because observations of SFXTs during their active outburst phases are fairly limited,
theoretical models for these accretion regimes
have not been well established and remain under debate.
Further MAXI-NICER coordinated operations 
as well as new time-domain X-ray missions
for capturing this type of fast X-ray transients,
such as Einstein-Probe X-ray satellite \citep{2022hxga.book...86Y},
will help resolve these problem.

\begin{acknowledgments}
MS acknowledges support from the Chinese Academy of
Sciences (CAS) President's International Fellowship Initiative (PIFI)
(grant No. 2020FSM004).
\end{acknowledgments}

\facilities{NICER(XTI), MAXI(GSC)}

\software{
  Astropy \citep{astropy2013,astropy2018},
  Matplotlib \citep{Hunter:2007},
  HEASoft \citep[v6.32.1][]{2014ascl.soft08004N},
  Stingray \citep{2019ApJ...881...39H,Huppenkothen2019},
  Xspec \citep[v12.13.1][]{arn96},
}

\appendix
\section{Grid scan observation}
\label{sec:gridscan}

\begin{figure}
\begin{center}
\includegraphics[width=1.\textwidth]{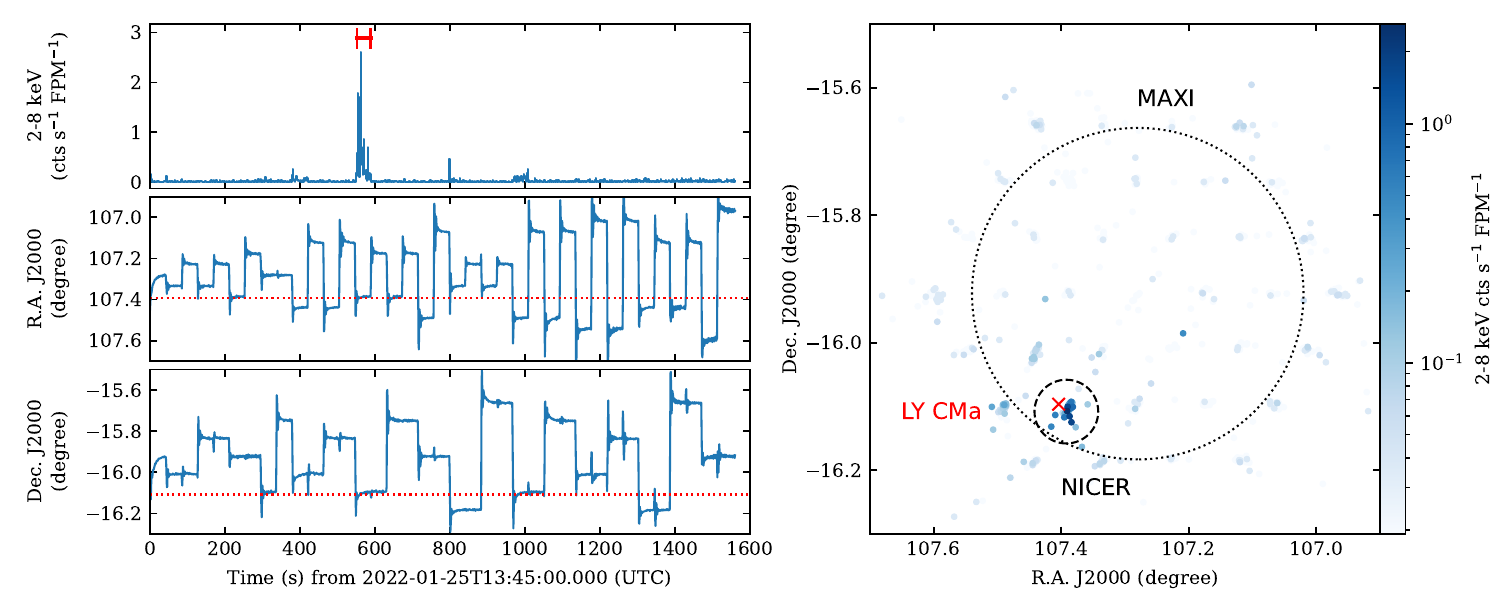}
\caption{
{\bf(Left)}
Time evolutions of NICER 2-8 kev count rate (top),
J2000 right ascension (middle)
and declination (bottom) of the XTI pointing direction
during the grid scan 
of the MAXI J0709 position-error region determined by MAXI. 
In the top panel,
the good time interval determined from the pointing direction 
is presented in a red line.
In the middle and bottom panels, dotted lines correspond to
the refined target coordinates.
{\bf(Right)}
Count-rate map in the scanned region. 
A dotted line represents the error circle  
of a $0\fdg26$ radius (systematic $\simeq0\fdg1$ plus statistic $\simeq0\fdg16$)
given by MAXI.
A dashed line represents the NICER error circle 
of a $3\arcmin$ radius. 
The data within the NICER circle was used in the source analysis in section \ref{sec:analysis}.
A red-cross mark indicates the position of the optical counterpart, LY CMa,
identified by the NuSTAR observation.
}
\label{fig:gridscan}
\end{center}
\end{figure}

The first $\sim 30$ minutes of the NICER observation 
was dedicate to the grid scan covering the entire error region
reported by MAXI \citep{serino2022ATel}
in order to confirm the new X-ray source
and determine the position with the better accuracy.
In figure \ref{fig:gridscan},
time evolutions of the 2--8 keV count rate,
and the NICER/XTI pointing direction
during the observation
are plotted on the left three panels 
and the count rate map on the scanned sky area
is plotted on the right panel.
The count rate increased significantly over the background level
when the XTI was pointed towards a certain direction \citep{iwakiri2022ATel}
within a radius of $\sim3\arcmin$, which is consistent with the size of the XTI field of view.
We defined the good time interval for which the XTI was pointed towards the region
within the $3\arcmin$
as shown on the count-rate evolution in figure \ref{fig:gridscan}
and used it in selecting target events in the following analysis,
as described in section \ref{sec:observation}.
The source position was later refined 
by the NuSTAR observation
and the object was identified to
the optical counterpart LY CMa
\citep[][]{sugizaki2022},
as plotted on the count-rate map in figure \ref{fig:gridscan}.

\bibliography{ms_maxij0709ni2}{}

\begin{thebibliography}{}
\expandafter\ifx\csname natexlab\endcsname\relax\def\natexlab#1{#1}\fi
\providecommand{\url}[1]{\href{#1}{#1}}
\providecommand{\dodoi}[1]{doi:~\href{http://doi.org/#1}{\nolinkurl{#1}}}
\providecommand{\doeprint}[1]{\href{http://ascl.net/#1}{\nolinkurl{http://ascl.net/#1}}}
\providecommand{\doarXiv}[1]{\href{https://arxiv.org/abs/#1}{\nolinkurl{https://arxiv.org/abs/#1}}}

\bibitem[{{Alpar} \& {Shaham}(1985)}]{1985Natur.316..239A}
{Alpar}, M.~A., \& {Shaham}, J. 1985, \nat, 316, 239, \dodoi{10.1038/316239a0}

\bibitem[{{Angelini} {et~al.}(1989){Angelini}, {Stella}, \&
  {Parmar}}]{1989ApJ...346..906A}
{Angelini}, L., {Stella}, L., \& {Parmar}, A.~N. 1989, \apj, 346, 906,
  \dodoi{10.1086/168070}

\bibitem[{{Arnaud}(1996)}]{arn96}
{Arnaud}, K.~A. 1996, in Astronomical Society of the Pacific Conference Series,
  Vol. 101, Astronomical Data Analysis Software and Systems V, ed. G.~H.
  {Jacoby} \& J.~{Barnes}, 17

\bibitem[{{Astropy Collaboration} {et~al.}(2013){Astropy Collaboration},
  {Robitaille}, {Tollerud}, {Greenfield}, {Droettboom}, {Bray}, {Aldcroft},
  {Davis}, {Ginsburg}, {Price-Whelan}, {Kerzendorf}, {Conley}, {Crighton},
  {Barbary}, {Muna}, {Ferguson}, {Grollier}, {Parikh}, {Nair}, {Unther},
  {Deil}, {Woillez}, {Conseil}, {Kramer}, {Turner}, {Singer}, {Fox}, {Weaver},
  {Zabalza}, {Edwards}, {Azalee Bostroem}, {Burke}, {Casey}, {Crawford},
  {Dencheva}, {Ely}, {Jenness}, {Labrie}, {Lim}, {Pierfederici}, {Pontzen},
  {Ptak}, {Refsdal}, {Servillat}, \& {Streicher}}]{astropy2013}
{Astropy Collaboration}, {Robitaille}, T.~P., {Tollerud}, E.~J., {et~al.} 2013,
  \aap, 558, A33, \dodoi{10.1051/0004-6361/201322068}

\bibitem[{{Astropy Collaboration} {et~al.}(2018){Astropy Collaboration},
  {Price-Whelan}, {Sip{\H{o}}cz}, {G{\"u}nther}, {Lim}, {Crawford}, {Conseil},
  {Shupe}, {Craig}, {Dencheva}, {Ginsburg}, {VanderPlas}, {Bradley},
  {P{\'e}rez-Su{\'a}rez}, {de Val-Borro}, {Aldcroft}, {Cruz}, {Robitaille},
  {Tollerud}, {Ardelean}, {Babej}, {Bach}, {Bachetti}, {Bakanov}, {Bamford},
  {Barentsen}, {Barmby}, {Baumbach}, {Berry}, {Biscani}, {Boquien}, {Bostroem},
  {Bouma}, {Brammer}, {Bray}, {Breytenbach}, {Buddelmeijer}, {Burke},
  {Calderone}, {Cano Rodr{\'\i}guez}, {Cara}, {Cardoso}, {Cheedella}, {Copin},
  {Corrales}, {Crichton}, {D'Avella}, {Deil}, {Depagne}, {Dietrich}, {Donath},
  {Droettboom}, {Earl}, {Erben}, {Fabbro}, {Ferreira}, {Finethy}, {Fox},
  {Garrison}, {Gibbons}, {Goldstein}, {Gommers}, {Greco}, {Greenfield},
  {Groener}, {Grollier}, {Hagen}, {Hirst}, {Homeier}, {Horton}, {Hosseinzadeh},
  {Hu}, {Hunkeler}, {Ivezi{\'c}}, {Jain}, {Jenness}, {Kanarek}, {Kendrew},
  {Kern}, {Kerzendorf}, {Khvalko}, {King}, {Kirkby}, {Kulkarni}, {Kumar},
  {Lee}, {Lenz}, {Littlefair}, {Ma}, {Macleod}, {Mastropietro}, {McCully},
  {Montagnac}, {Morris}, {Mueller}, {Mumford}, {Muna}, {Murphy}, {Nelson},
  {Nguyen}, {Ninan}, {N{\"o}the}, {Ogaz}, {Oh}, {Parejko}, {Parley}, {Pascual},
  {Patil}, {Patil}, {Plunkett}, {Prochaska}, {Rastogi}, {Reddy Janga},
  {Sabater}, {Sakurikar}, {Seifert}, {Sherbert}, {Sherwood-Taylor}, {Shih},
  {Sick}, {Silbiger}, {Singanamalla}, {Singer}, {Sladen}, {Sooley},
  {Sornarajah}, {Streicher}, {Teuben}, {Thomas}, {Tremblay}, {Turner},
  {Terr{\'o}n}, {van Kerkwijk}, {de la Vega}, {Watkins}, {Weaver}, {Whitmore},
  {Woillez}, {Zabalza}, \& {Astropy Contributors}}]{astropy2018}
{Astropy Collaboration}, {Price-Whelan}, A.~M., {Sip{\H{o}}cz}, B.~M., {et~al.}
  2018, \aj, 156, 123, \dodoi{10.3847/1538-3881/aabc4f}

\bibitem[{{Bailer-Jones} {et~al.}(2021){Bailer-Jones}, {Rybizki}, {Fouesneau},
  {Demleitner}, \& {Andrae}}]{2021AJ....161..147B}
{Bailer-Jones}, C.~A.~L., {Rybizki}, J., {Fouesneau}, M., {Demleitner}, M., \&
  {Andrae}, R. 2021, \aj, 161, 147, \dodoi{10.3847/1538-3881/abd806}

\bibitem[{{Belloni} {et~al.}(2002){Belloni}, {Psaltis}, \& {van der
  Klis}}]{2002ApJ...572..392B}
{Belloni}, T., {Psaltis}, D., \& {van der Klis}, M. 2002, \apj, 572, 392,
  \dodoi{10.1086/340290}

\bibitem[{{Bhattacharya} {et~al.}(2024){Bhattacharya}, {Mathew}, {Banerjee},
  {G}, {Muneer}, {Pramod Kumar}, \& {Joshi}}]{2024BSRSL..93..636B}
{Bhattacharya}, S., {Mathew}, B., {Banerjee}, G., {et~al.} 2024, Bulletin de la
  Societe Royale des Sciences de Liege, 93, 636,
  \dodoi{10.25518/0037-9565.11821}

\bibitem[{{Bhattacharyya} {et~al.}(2022){Bhattacharyya}, {Mathew}, {Ezhikode},
  {Muneer}, {Selvakumar}, {Maheswer}, {Arun}, {Anilkumar}, {Banerjee},
  {Pramod}, {Kartha}, {Paul}, \& {Velu}}]{bhattacharyya2022}
{Bhattacharyya}, S., {Mathew}, B., {Ezhikode}, S.~H., {et~al.} 2022, \apjl,
  933, L34, \dodoi{10.3847/2041-8213/ac7b8a}

\bibitem[{{Bondi}(1952)}]{1952MNRAS.112..195B}
{Bondi}, H. 1952, \mnras, 112, 195, \dodoi{10.1093/mnras/112.2.195}

\bibitem[{{Bozzo} {et~al.}(2017){Bozzo}, {Bernardini}, {Ferrigno}, {Falanga},
  {Romano}, \& {Oskinova}}]{2017A&A...608A.128B}
{Bozzo}, E., {Bernardini}, F., {Ferrigno}, C., {et~al.} 2017, \aap, 608, A128,
  \dodoi{10.1051/0004-6361/201730398}

\bibitem[{{Bozzo} {et~al.}(2008){Bozzo}, {Falanga}, \&
  {Stella}}]{2008ApJ...683.1031B}
{Bozzo}, E., {Falanga}, M., \& {Stella}, L. 2008, \apj, 683, 1031,
  \dodoi{10.1086/589990}

\bibitem[{{Bozzo} {et~al.}(2016{\natexlab{a}}){Bozzo}, {Oskinova}, {Feldmeier},
  \& {Falanga}}]{2016A&A...589A.102B}
{Bozzo}, E., {Oskinova}, L., {Feldmeier}, A., \& {Falanga}, M.
  2016{\natexlab{a}}, \aap, 589, A102, \dodoi{10.1051/0004-6361/201628341}

\bibitem[{{Bozzo} {et~al.}(2015){Bozzo}, {Romano}, {Ducci}, {Bernardini}, \&
  {Falanga}}]{2015AdSpR..55.1255B}
{Bozzo}, E., {Romano}, P., {Ducci}, L., {Bernardini}, F., \& {Falanga}, M.
  2015, Advances in Space Research, 55, 1255, \dodoi{10.1016/j.asr.2014.11.012}

\bibitem[{{Bozzo} {et~al.}(2009){Bozzo}, {Stella}, {Vietri}, \&
  {Ghosh}}]{2009A&A...493..809B}
{Bozzo}, E., {Stella}, L., {Vietri}, M., \& {Ghosh}, P. 2009, \aap, 493, 809,
  \dodoi{10.1051/0004-6361:200810658}

\bibitem[{{Bozzo} {et~al.}(2011){Bozzo}, {Giunta}, {Cusumano}, {Ferrigno},
  {Walter}, {Campana}, {Falanga}, {Israel}, \& {Stella}}]{2011A&A...531A.130B}
{Bozzo}, E., {Giunta}, A., {Cusumano}, G., {et~al.} 2011, \aap, 531, A130,
  \dodoi{10.1051/0004-6361/201116726}

\bibitem[{{Bozzo} {et~al.}(2016{\natexlab{b}}){Bozzo}, {Bhalerao}, {Pradhan},
  {Tomsick}, {Romano}, {Ferrigno}, {Chaty}, {Oskinova}, {Manousakis}, {Walter},
  {Falanga}, {Campana}, {Stella}, {Ramolla}, \& {Chini}}]{2016A&A...596A..16B}
{Bozzo}, E., {Bhalerao}, V., {Pradhan}, P., {et~al.} 2016{\natexlab{b}}, \aap,
  596, A16, \dodoi{10.1051/0004-6361/201629311}

\bibitem[{{Cash}(1979)}]{Cash1979}
{Cash}, W. 1979, \apj, 228, 939, \dodoi{10.1086/156922}

\bibitem[{{Castor} {et~al.}(1975){Castor}, {Abbott}, \&
  {Klein}}]{1975ApJ...195..157C}
{Castor}, J.~I., {Abbott}, D.~C., \& {Klein}, R.~I. 1975, \apj, 195, 157,
  \dodoi{10.1086/153315}

\bibitem[{{Chojnowski} {et~al.}(2015){Chojnowski}, {Whelan}, {Wisniewski},
  {Majewski}, {Hall}, {Shetrone}, {Beaton}, {Burton}, {Damke}, {Eikenberry},
  {Hasselquist}, {Holtzman}, {M{\'e}sz{\'a}ros}, {Nidever}, {Schneider},
  {Wilson}, {Zasowski}, {Bizyaev}, {Brewington}, {Brinkmann}, {Ebelke},
  {Frinchaboy}, {Kinemuchi}, {Malanushenko}, {Malanushenko}, {Marchante},
  {Oravetz}, {Pan}, \& {Simmons}}]{2015AJ....149....7C}
{Chojnowski}, S.~D., {Whelan}, D.~G., {Wisniewski}, J.~P., {et~al.} 2015, \aj,
  149, 7, \dodoi{10.1088/0004-6256/149/1/7}

\bibitem[{{Coburn} {et~al.}(2002){Coburn}, {Heindl}, {Rothschild}, {Gruber},
  {Kreykenbohm}, {Wilms}, {Kretschmar}, \& {Staubert}}]{2002ApJ...580..394C}
{Coburn}, W., {Heindl}, W.~A., {Rothschild}, R.~E., {et~al.} 2002, \apj, 580,
  394, \dodoi{10.1086/343033}

\bibitem[{{Crowther} {et~al.}(2006){Crowther}, {Lennon}, \&
  {Walborn}}]{2006A&A...446..279C}
{Crowther}, P.~A., {Lennon}, D.~J., \& {Walborn}, N.~R. 2006, \aap, 446, 279,
  \dodoi{10.1051/0004-6361:20053685}

\bibitem[{{Devasia} {et~al.}(2011){Devasia}, {James}, {Paul}, \&
  {Indulekha}}]{2011MNRAS.417..348D}
{Devasia}, J., {James}, M., {Paul}, B., \& {Indulekha}, K. 2011, \mnras, 417,
  348, \dodoi{10.1111/j.1365-2966.2011.19269.x}

\bibitem[{{Di Salvo} \& {Sanna}(2020)}]{2020arXiv201009005D}
{Di Salvo}, T., \& {Sanna}, A. 2020, arXiv e-prints, arXiv:2010.09005,
  \dodoi{10.48550/arXiv.2010.09005}

\bibitem[{{Ducci} {et~al.}(2010){Ducci}, {Sidoli}, \&
  {Paizis}}]{2010MNRAS.408.1540D}
{Ducci}, L., {Sidoli}, L., \& {Paizis}, A. 2010, \mnras, 408, 1540,
  \dodoi{10.1111/j.1365-2966.2010.17216.x}

\bibitem[{{Finger} {et~al.}(1996){Finger}, {Wilson}, \&
  {Harmon}}]{1996ApJ...459..288F}
{Finger}, M.~H., {Wilson}, R.~B., \& {Harmon}, B.~A. 1996, \apj, 459, 288,
  \dodoi{10.1086/176892}

\bibitem[{{Gendreau} {et~al.}(2016){Gendreau}, {Arzoumanian}, {Adkins},
  {Albert}, {Anders}, {Aylward}, {Baker}, {Balsamo}, {Bamford}, {Benegalrao},
  {Berry}, {Bhalwani}, {Black}, {Blaurock}, {Bronke}, {Brown}, {Budinoff},
  {Cantwell}, {Cazeau}, {Chen}, {Clement}, {Colangelo}, {Coleman},
  {Coopersmith}, {Dehaven}, {Doty}, {Egan}, {Enoto}, {Fan}, {Ferro}, {Foster},
  {Galassi}, {Gallo}, {Green}, {Grosh}, {Ha}, {Hasouneh}, {Heefner}, {Hestnes},
  {Hoge}, {Jacobs}, {J{\o}rgensen}, {Kaiser}, {Kellogg}, {Kenyon}, {Koenecke},
  {Kozon}, {LaMarr}, {Lambertson}, {Larson}, {Lentine}, {Lewis}, {Lilly},
  {Liu}, {Malonis}, {Manthripragada}, {Markwardt}, {Matonak}, {Mcginnis},
  {Miller}, {Mitchell}, {Mitchell}, {Mohammed}, {Monroe}, {Montt de Garcia},
  {Mul{\'e}}, {Nagao}, {Ngo}, {Norris}, {Norwood}, {Novotka}, {Okajima},
  {Olsen}, {Onyeachu}, {Orosco}, {Peterson}, {Pevear}, {Pham}, {Pollard},
  {Pope}, {Powers}, {Powers}, {Price}, {Prigozhin}, {Ramirez}, {Reid},
  {Remillard}, {Rogstad}, {Rosecrans}, {Rowe}, {Sager}, {Sanders}, {Savadkin},
  {Saylor}, {Schaeffer}, {Schweiss}, {Semper}, {Serlemitsos}, {Shackelford},
  {Soong}, {Struebel}, {Vezie}, {Villasenor}, {Winternitz}, {Wofford},
  {Wright}, {Yang}, \& {Yu}}]{2016SPIE.9905E..1HG}
{Gendreau}, K.~C., {Arzoumanian}, Z., {Adkins}, P.~W., {et~al.} 2016, in
  Society of Photo-Optical Instrumentation Engineers (SPIE) Conference Series,
  Vol. 9905, Space Telescopes and Instrumentation 2016: Ultraviolet to Gamma
  Ray, ed. J.-W.~A. {den Herder}, T.~{Takahashi}, \& M.~{Bautz}, 99051H,
  \dodoi{10.1117/12.2231304}

\bibitem[{{Ghosh} \& {Lamb}(1979{\natexlab{a}})}]{GhoshLamb1979II}
{Ghosh}, P., \& {Lamb}, F.~K. 1979{\natexlab{a}}, \apj, 232, 259,
  \dodoi{10.1086/157285}

\bibitem[{{Ghosh} \& {Lamb}(1979{\natexlab{b}})}]{GhoshLamb1979III}
---. 1979{\natexlab{b}}, \apj, 234, 296, \dodoi{10.1086/157498}

\bibitem[{{Gim{\'e}nez-Garc{\'\i}a} {et~al.}(2015){Gim{\'e}nez-Garc{\'\i}a},
  {Torrej{\'o}n}, {Eikmann}, {Mart{\'\i}nez-N{\'u}{\~n}ez}, {Oskinova},
  {Rodes-Roca}, \& {Bernab{\'e}u}}]{2015A&A...576A.108G}
{Gim{\'e}nez-Garc{\'\i}a}, A., {Torrej{\'o}n}, J.~M., {Eikmann}, W., {et~al.}
  2015, \aap, 576, A108, \dodoi{10.1051/0004-6361/201425004}

\bibitem[{{Gonz{\'a}lez-Gal{\'a}n} {et~al.}(2014){Gonz{\'a}lez-Gal{\'a}n},
  {Negueruela}, {Castro}, {Sim{\'o}n-D{\'\i}az}, {Lorenzo}, \&
  {Vilardell}}]{2014A&A...566A.131G}
{Gonz{\'a}lez-Gal{\'a}n}, A., {Negueruela}, I., {Castro}, N., {et~al.} 2014,
  \aap, 566, A131, \dodoi{10.1051/0004-6361/201423554}

\bibitem[{{Grebenev} \& {Sunyaev}(2007)}]{2007AstL...33..149G}
{Grebenev}, S.~A., \& {Sunyaev}, R.~A. 2007, Astronomy Letters, 33, 149,
  \dodoi{10.1134/S1063773707030024}

\bibitem[{{Hainich} {et~al.}(2020){Hainich}, {Oskinova}, {Torrej{\'o}n},
  {Fuerst}, {Bodaghee}, {Shenar}, {Sander}, {Todt}, {Spetzer}, \&
  {Hamann}}]{2020A&A...634A..49H}
{Hainich}, R., {Oskinova}, L.~M., {Torrej{\'o}n}, J.~M., {et~al.} 2020, \aap,
  634, A49, \dodoi{10.1051/0004-6361/201935498}

\bibitem[{{Harrison} {et~al.}(2013){Harrison}, {Craig}, {Christensen},
  {Hailey}, {Zhang}, {Boggs}, {Stern}, {Cook}, {Forster}, {Giommi},
  {Grefenstette}, {Kim}, {Kitaguchi}, {Koglin}, {Madsen}, {Mao}, {Miyasaka},
  {Mori}, {Perri}, {Pivovaroff}, {Puccetti}, {Rana}, {Westergaard}, {Willis},
  {Zoglauer}, {An}, {Bachetti}, {Barri{\`e}re}, {Bellm}, {Bhalerao},
  {Brejnholt}, {Fuerst}, {Liebe}, {Markwardt}, {Nynka}, {Vogel}, {Walton},
  {Wik}, {Alexander}, {Cominsky}, {Hornschemeier}, {Hornstrup}, {Kaspi},
  {Madejski}, {Matt}, {Molendi}, {Smith}, {Tomsick}, {Ajello}, {Ballantyne},
  {Balokovi{\'c}}, {Barret}, {Bauer}, {Blandford}, {Brandt}, {Brenneman},
  {Chiang}, {Chakrabarty}, {Chenevez}, {Comastri}, {Dufour}, {Elvis}, {Fabian},
  {Farrah}, {Fryer}, {Gotthelf}, {Grindlay}, {Helfand}, {Krivonos}, {Meier},
  {Miller}, {Natalucci}, {Ogle}, {Ofek}, {Ptak}, {Reynolds}, {Rigby},
  {Tagliaferri}, {Thorsett}, {Treister}, \& {Urry}}]{harrison2013}
{Harrison}, F.~A., {Craig}, W.~W., {Christensen}, F.~E., {et~al.} 2013, \apj,
  770, 103, \dodoi{10.1088/0004-637X/770/2/103}

\bibitem[{{HI4PI Collaboration} {et~al.}(2016){HI4PI Collaboration}, {Ben
  Bekhti}, {Fl{\"o}er}, {Keller}, {Kerp}, {Lenz}, {Winkel}, {Bailin},
  {Calabretta}, {Dedes}, {Ford}, {Gibson}, {Haud}, {Janowiecki}, {Kalberla},
  {Lockman}, {McClure-Griffiths}, {Murphy}, {Nakanishi}, {Pisano}, \&
  {Staveley-Smith}}]{hi416}
{HI4PI Collaboration}, {Ben Bekhti}, N., {Fl{\"o}er}, L., {et~al.} 2016, \aap,
  594, A116, \dodoi{10.1051/0004-6361/201629178}

\bibitem[{{Houk} \& {Smith-Moore}(1988)}]{1988mcts.book.....H}
{Houk}, N., \& {Smith-Moore}, M. 1988, {Michigan Catalogue of Two-dimensional
  Spectral Types for the HD Stars. Volume 4, Declinations -26{\textdegree}.0 to
  -12{\textdegree}.0.}, Vol.~4

\bibitem[{Hunter(2007)}]{Hunter:2007}
Hunter, J.~D. 2007, Computing in Science \& Engineering, 9, 90,
  \dodoi{10.1109/MCSE.2007.55}

\bibitem[{{Huppenkothen} {et~al.}(2019{\natexlab{a}}){Huppenkothen},
  {Bachetti}, {Stevens}, {Migliari}, {Balm}, {Hammad}, {Khan}, {Mishra},
  {Rashid}, {Sharma}, {Martinez Ribeiro}, \& {Valles
  Blanco}}]{2019ApJ...881...39H}
{Huppenkothen}, D., {Bachetti}, M., {Stevens}, A.~L., {et~al.}
  2019{\natexlab{a}}, \apj, 881, 39, \dodoi{10.3847/1538-4357/ab258d}

\bibitem[{{Huppenkothen} {et~al.}(2019{\natexlab{b}}){Huppenkothen},
  {Bachetti}, {Stevens}, {Migliari}, {Balm}, {Hammad}, {Khan}, {Mishra},
  {Rashid}, {Sharma}, {Martinez Ribeiro}, \& {Valles
  Blanco}}]{Huppenkothen2019}
---. 2019{\natexlab{b}}, Journal of Open Source Software, 4, 1393,
  \dodoi{10.21105/joss.01393}

\bibitem[{{Inoue}(1985)}]{1985SSRv...40..317I}
{Inoue}, H. 1985, \ssr, 40, 317, \dodoi{10.1007/BF00212905}

\bibitem[{{Iwakiri} {et~al.}(2022){Iwakiri}, {Gendreau}, {Arzoumanian},
  {Enoto}, {Mihara}, {Serino}, {Negoro}, {Pope}, \& {Sanna}}]{iwakiri2022ATel}
{Iwakiri}, W., {Gendreau}, K., {Arzoumanian}, Z., {et~al.} 2022, The
  Astronomer's Telegram, 15181, 1

\bibitem[{{James} {et~al.}(2010){James}, {Paul}, {Devasia}, \&
  {Indulekha}}]{2010MNRAS.407..285J}
{James}, M., {Paul}, B., {Devasia}, J., \& {Indulekha}, K. 2010, \mnras, 407,
  285, \dodoi{10.1111/j.1365-2966.2010.16880.x}

\bibitem[{{Kaur} {et~al.}(2007){Kaur}, {Paul}, {Raichur}, \&
  {Sagar}}]{2007ApJ...660.1409K}
{Kaur}, R., {Paul}, B., {Raichur}, H., \& {Sagar}, R. 2007, \apj, 660, 1409,
  \dodoi{10.1086/513418}

\bibitem[{{Kobayashi} {et~al.}(2022){Kobayashi}, {Negoro}, {Sugita}, {Serino},
  {Mihara}, {Iwakiri}, {Nakajima}, {Asakura}, {Seino}, {Tamagawa}, {Li},
  {Matsuoka}, {Sakamoto}, {Komachi}, {Hiramatsu}, {Yoshida}, {Tsuboi}, {Kawai},
  {Okamoto}, {Kitakoga}, {Kohara}, {Shidatsu}, {Iwasaki}, {Kawai}, {Niwano},
  {Hosokawa}, {Imai}, {Ito}, {Takamatsu}, {Nakahira}, {Ueno}, {Tomida},
  {Ishikawa}, {Tominaga}, {Nagatsuka}, {Kurihara}, {Ueda}, {Yamada}, {Ogawa},
  {Setoguchi}, {Yoshitake}, {Goto}, {Uematsu}, {Inaba}, {Tsunemi}, {Yamauchi},
  {Nonaka}, {Sato}, {Hatsuda}, {Fukuoka}, {Kawamuro}, {Yamaoka}, {Kawakubo}, \&
  {Sugizaki}}]{kobayashi2022ATel}
{Kobayashi}, K., {Negoro}, H., {Sugita}, S., {et~al.} 2022, The Astronomer's
  Telegram, 15188, 1

\bibitem[{{Kretschmar} {et~al.}(2019){Kretschmar}, {F{\"u}rst}, {Sidoli},
  {Bozzo}, {Alfonso-Garz{\'o}n}, {Bodaghee}, {Chaty}, {Chernyakova},
  {Ferrigno}, {Manousakis}, {Negueruela}, {Postnov}, {Paizis}, {Reig},
  {Rodes-Roca}, {Tsygankov}, {Bird}, {Bissinger n{\'e} K{\"u}hnel}, {Blay},
  {Caballero}, {Coe}, {Domingo}, {Doroshenko}, {Ducci}, {Falanga}, {Grebenev},
  {Grinberg}, {Hemphill}, {Kreykenbohm}, {Kreykenbohm n{\'e} Fritz}, {Li},
  {Lutovinov}, {Mart{\'\i}nez-N{\'u}{\~n}ez}, {Mas-Hesse}, {Masetti},
  {McBride}, {Neronov}, {Pottschmidt}, {Rodriguez}, {Romano}, {Rothschild},
  {Santangelo}, {Sguera}, {Staubert}, {Tomsick}, {Torrej{\'o}n}, {Torres},
  {Walter}, {Wilms}, {Wilson-Hodge}, \& {Zhang}}]{2019NewAR..8601546K}
{Kretschmar}, P., {F{\"u}rst}, F., {Sidoli}, L., {et~al.} 2019, \nar, 86,
  101546, \dodoi{10.1016/j.newar.2020.101546}

\bibitem[{{Kreykenbohm} {et~al.}(2008){Kreykenbohm}, {Wilms}, {Kretschmar},
  {Torrej{\'o}n}, {Pottschmidt}, {Hanke}, {Santangelo}, {Ferrigno}, \&
  {Staubert}}]{2008A&A...492..511K}
{Kreykenbohm}, I., {Wilms}, J., {Kretschmar}, P., {et~al.} 2008, \aap, 492,
  511, \dodoi{10.1051/0004-6361:200809956}

\bibitem[{{Krti{\v{c}}ka} \& {Kub{\'a}t}(2011)}]{2011A&A...534A..97K}
{Krti{\v{c}}ka}, J., \& {Kub{\'a}t}, J. 2011, \aap, 534, A97,
  \dodoi{10.1051/0004-6361/201116679}

\bibitem[{{Leahy} {et~al.}(1983){Leahy}, {Darbro}, {Elsner}, {Weisskopf},
  {Sutherland}, {Kahn}, \& {Grindlay}}]{1983ApJ...266..160L}
{Leahy}, D.~A., {Darbro}, W., {Elsner}, R.~F., {et~al.} 1983, \apj, 266, 160,
  \dodoi{10.1086/160766}

\bibitem[{{Liu} {et~al.}(2022){Liu}, {Wang}, {Chen}, {Yang}, {Lu}, {Song},
  {Qu}, {Zhang}, \& {Zhang}}]{2022MNRAS.516.5579L}
{Liu}, Q., {Wang}, W., {Chen}, X., {et~al.} 2022, \mnras, 516, 5579,
  \dodoi{10.1093/mnras/stac2646}

\bibitem[{{Makishima} {et~al.}(1999){Makishima}, {Mihara}, {Nagase}, \&
  {Tanaka}}]{1999ApJ...525..978M}
{Makishima}, K., {Mihara}, T., {Nagase}, F., \& {Tanaka}, Y. 1999, \apj, 525,
  978, \dodoi{10.1086/307912}

\bibitem[{{Mart{\'\i}nez-N{\'u}{\~n}ez}
  {et~al.}(2017){Mart{\'\i}nez-N{\'u}{\~n}ez}, {Kretschmar}, {Bozzo},
  {Oskinova}, {Puls}, {Sidoli}, {Sundqvist}, {Blay}, {Falanga}, {F{\"u}rst},
  {G{\'\i}menez-Garc{\'\i}a}, {Kreykenbohm}, {K{\"u}hnel}, {Sander},
  {Torrej{\'o}n}, \& {Wilms}}]{2017SSRv..212...59M}
{Mart{\'\i}nez-N{\'u}{\~n}ez}, S., {Kretschmar}, P., {Bozzo}, E., {et~al.}
  2017, \ssr, 212, 59, \dodoi{10.1007/s11214-017-0340-1}

\bibitem[{{Matsuoka} {et~al.}(2009){Matsuoka}, {Kawasaki}, {Ueno}, {Tomida},
  {Kohama}, {Suzuki}, {Adachi}, {Ishikawa}, {Mihara}, {Sugizaki}, {Isobe},
  {Nakagawa}, {Tsunemi}, {Miyata}, {Kawai}, {Kataoka}, {Morii}, {Yoshida},
  {Negoro}, {Nakajima}, {Ueda}, {Chujo}, {Yamaoka}, {Yamazaki}, {Nakahira},
  {You}, {Ishiwata}, {Miyoshi}, {Eguchi}, {Hiroi}, {Katayama}, \&
  {Ebisawa}}]{matsuoka09}
{Matsuoka}, M., {Kawasaki}, K., {Ueno}, S., {et~al.} 2009, \pasj, 61, 999,
  \dodoi{10.1093/pasj/61.5.999}

\bibitem[{{Mihara} {et~al.}(2011){Mihara}, {Nakajima}, {Sugizaki}, {Serino},
  {Matsuoka}, {Kohama}, {Kawasaki}, {Tomida}, {Ueno}, {Kawai}, {Kataoka},
  {Morii}, {Yoshida}, {Yamaoka}, {Nakahira}, {Negoro}, {Isobe}, {Yamauchi}, \&
  {Sakurai}}]{2011PASJ...63S.623M}
{Mihara}, T., {Nakajima}, M., {Sugizaki}, M., {et~al.} 2011, \pasj, 63, S623,
  \dodoi{10.1093/pasj/63.sp3.S623}

\bibitem[{{NASA High Energy Astrophysics Science Archive Research Center
  (HEASARC)}(2014)}]{2014ascl.soft08004N}
{NASA High Energy Astrophysics Science Archive Research Center (HEASARC)}.
  2014, {HEAsoft: Unified Release of FTOOLS and XANADU}, Astrophysics Source
  Code Library, record ascl:1408.004.
\newblock \doeprint{1408.004}

\bibitem[{{Negoro} {et~al.}(2022){Negoro}, {Mihara}, {Pike}, {Grefenstette},
  {Iwakiri}, {Nakajima}, \& {Gendreau}}]{negoro2022ATel}
{Negoro}, H., {Mihara}, T., {Pike}, S., {et~al.} 2022, The Astronomer's
  Telegram, 15193, 1

\bibitem[{{Nesci}(2022)}]{nesci2022ATel}
{Nesci}, R. 2022, The Astronomer's Telegram, 15194, 1

\bibitem[{{Pike} {et~al.}(2023){Pike}, {Sugizaki}, {Eijnden}, {Coughenour},
  {Jaodand}, {Mihara}, {Motta}, {Negoro}, {Shaw}, {Shidatsu}, \&
  {Tomsick}}]{2023ApJ...954...48P}
{Pike}, S.~N., {Sugizaki}, M., {Eijnden}, J. v.~d., {et~al.} 2023, \apj, 954,
  48, \dodoi{10.3847/1538-4357/ace696}

\bibitem[{{Pradhan} {et~al.}(2018){Pradhan}, {Bozzo}, \&
  {Paul}}]{2018A&A...610A..50P}
{Pradhan}, P., {Bozzo}, E., \& {Paul}, B. 2018, \aap, 610, A50,
  \dodoi{10.1051/0004-6361/201731487}

\bibitem[{{Prinja}(1989)}]{1989MNRAS.241..721P}
{Prinja}, R.~K. 1989, \mnras, 241, 721, \dodoi{10.1093/mnras/241.4.721}

\bibitem[{{Rahoui} \& {Chaty}(2008)}]{2008A&A...492..163R}
{Rahoui}, F., \& {Chaty}, S. 2008, \aap, 492, 163,
  \dodoi{10.1051/0004-6361:200810695}

\bibitem[{{Reig}(2011)}]{2011Ap&SS.332....1R}
{Reig}, P. 2011, \apss, 332, 1, \dodoi{10.1007/s10509-010-0575-8}

\bibitem[{{Rikame} {et~al.}(2022){Rikame}, {Paul}, {Pradhan}, \&
  {Paul}}]{2022MNRAS.512.4792R}
{Rikame}, K., {Paul}, B., {Pradhan}, P., \& {Paul}, K.~T. 2022, \mnras, 512,
  4792, \dodoi{10.1093/mnras/stac729}

\bibitem[{{Romano} {et~al.}(2013){Romano}, {Mangano}, {Ducci}, {Esposito},
  {Vercellone}, {Bocchino}, {Burrows}, {Kennea}, {Krimm}, {Gehrels},
  {Farinelli}, \& {Ceccobello}}]{2013AdSpR..52.1593R}
{Romano}, P., {Mangano}, V., {Ducci}, L., {et~al.} 2013, Advances in Space
  Research, 52, 1593, \dodoi{10.1016/j.asr.2013.07.034}

\bibitem[{{Romano} {et~al.}(2015){Romano}, {Bozzo}, {Mangano}, {Esposito},
  {Israel}, {Tiengo}, {Campana}, {Ducci}, {Ferrigno}, \&
  {Kennea}}]{2015A&A...576L...4R}
{Romano}, P., {Bozzo}, E., {Mangano}, V., {et~al.} 2015, \aap, 576, L4,
  \dodoi{10.1051/0004-6361/201525749}

\bibitem[{{Roy} {et~al.}(2019){Roy}, {Agrawal}, {Iyer}, {Bhattacharya},
  {Yadav}, {Antia}, {Chauhan}, {Choudhury}, {Dedhia}, {Katoch}, {Madhavani},
  {Manchanda}, {Misra}, {Pahari}, {Paul}, \& {Shah}}]{2019ApJ...872...33R}
{Roy}, J., {Agrawal}, P.~C., {Iyer}, N.~K., {et~al.} 2019, \apj, 872, 33,
  \dodoi{10.3847/1538-4357/aafaf1}

\bibitem[{{Serino} {et~al.}(2022){Serino}, {Negoro}, {Nakajima}, {Kobayashi},
  {Asakura}, {Seino}, {Mihara}, {Tamagawa}, {Li}, {Matsuoka}, {Sakamoto},
  {Sugita}, {Komachi}, {Hiramatsu}, {Yoshida}, {Tsuboi}, {Iwakiri}, {Kawai},
  {Okamoto}, {Kitakoga}, {Kohara}, {Shidatsu}, {Iwasaki}, {Kawai}, {Niwano},
  {Hosokawa}, {Imai}, {Ito}, {Takamatsu}, {Nakahira}, {Ueno}, {Tomida},
  {Ishikawa}, {Tominaga}, {Nagatsuka}, {Kurihara}, {Ueda}, {Yamada}, {Ogawa},
  {Setoguchi}, {Yoshitake}, {Goto}, {Uematsu}, {Inaba}, {Tsunemi}, {Yamauchi},
  {Nonaka}, {Sato}, {Hatsuda}, {Fukuoka}, {Kawamuro}, {Yamaoka}, {Kawakubo}, \&
  {Sugizaki}}]{serino2022ATel}
{Serino}, M., {Negoro}, H., {Nakajima}, M., {et~al.} 2022, The Astronomer's
  Telegram, 15178, 1

\bibitem[{{Sguera} {et~al.}(2006){Sguera}, {Bazzano}, {Bird}, {Dean},
  {Ubertini}, {Barlow}, {Bassani}, {Clark}, {Hill}, {Malizia}, {Molina}, \&
  {Stephen}}]{2006ApJ...646..452S}
{Sguera}, V., {Bazzano}, A., {Bird}, A.~J., {et~al.} 2006, \apj, 646, 452,
  \dodoi{10.1086/504827}

\bibitem[{{Shakura} {et~al.}(2013){Shakura}, {Postnov}, \&
  {Hjalmarsdotter}}]{2013MNRAS.428..670S}
{Shakura}, N., {Postnov}, K., \& {Hjalmarsdotter}, L. 2013, \mnras, 428, 670,
  \dodoi{10.1093/mnras/sts062}

\bibitem[{{Shakura} {et~al.}(2014){Shakura}, {Postnov}, {Sidoli}, \&
  {Paizis}}]{2014MNRAS.442.2325S}
{Shakura}, N., {Postnov}, K., {Sidoli}, L., \& {Paizis}, A. 2014, \mnras, 442,
  2325, \dodoi{10.1093/mnras/stu1027}

\bibitem[{{Shidatsu} {et~al.}(2025){Shidatsu}, {Kawai}, {Maehara}, {Goto},
  {Urabe}, {Iwakiri}, {Tsuboi}, {Nemoto}, {Nawa}, {Sugizaki}, {Nakajima},
  {Niwano}, {Hosokawa}, {Sakamoto}, \& {Matsuoka}}]{2025arXiv250321118S}
{Shidatsu}, M., {Kawai}, N., {Maehara}, H., {et~al.} 2025, arXiv e-prints,
  arXiv:2503.21118, \dodoi{10.48550/arXiv.2503.21118}

\bibitem[{{Sidoli} {et~al.}(2016){Sidoli}, {Esposito}, {Motta}, {Israel}, \&
  {Rodr{\'\i}guez Castillo}}]{2016MNRAS.460.3637S}
{Sidoli}, L., {Esposito}, P., {Motta}, S.~E., {Israel}, G.~L., \&
  {Rodr{\'\i}guez Castillo}, G.~A. 2016, \mnras, 460, 3637,
  \dodoi{10.1093/mnras/stw1246}

\bibitem[{{Sidoli} \& {Paizis}(2018)}]{2018MNRAS.481.2779S}
{Sidoli}, L., \& {Paizis}, A. 2018, \mnras, 481, 2779,
  \dodoi{10.1093/mnras/sty2428}

\bibitem[{{Sugizaki} {et~al.}(2011){Sugizaki}, {Mihara}, {Serino}, {Yamamoto},
  {Matsuoka}, {Kohama}, {Tomida}, {Ueno}, {Kawai}, {Morii}, {Sugimori},
  {Nakahira}, {Yamaoka}, {Yoshida}, {Nakajima}, {Negoro}, {Eguchi}, {Isobe},
  {Ueda}, \& {Tsunemi}}]{2011PASJ...63S.635S}
{Sugizaki}, M., {Mihara}, T., {Serino}, M., {et~al.} 2011, \pasj, 63, S635,
  \dodoi{10.1093/pasj/63.sp3.S635}

\bibitem[{{Sugizaki} {et~al.}(2022){Sugizaki}, {Mihara}, {Kobayashi}, {Negoro},
  {Shidatsu}, {Pike}, {Iwakiri}, {Urabe}, {Serino}, {Kawai}, {Nakajima},
  {Kennea}, \& {Liu}}]{sugizaki2022}
{Sugizaki}, M., {Mihara}, T., {Kobayashi}, K., {et~al.} 2022, \pasj,
  \dodoi{10.1093/pasj/psac059}

\bibitem[{{Taam} {et~al.}(1988){Taam}, {Fryxell}, \&
  {Brown}}]{1988ApJ...331L.117T}
{Taam}, R.~E., {Fryxell}, B.~A., \& {Brown}, D.~A. 1988, \apjl, 331, L117,
  \dodoi{10.1086/185248}

\bibitem[{{Takeshima} {et~al.}(1994){Takeshima}, {Dotani}, {Mitsuda}, \&
  {Nagase}}]{1994ApJ...436..871T}
{Takeshima}, T., {Dotani}, T., {Mitsuda}, K., \& {Nagase}, F. 1994, \apj, 436,
  871, \dodoi{10.1086/174964}

\bibitem[{{Torrej{\'o}n} {et~al.}(2010){Torrej{\'o}n}, {Schulz}, {Nowak}, \&
  {Kallman}}]{2010ApJ...715..947T}
{Torrej{\'o}n}, J.~M., {Schulz}, N.~S., {Nowak}, M.~A., \& {Kallman}, T.~R.
  2010, \apj, 715, 947, \dodoi{10.1088/0004-637X/715/2/947}

\bibitem[{{van der Klis} {et~al.}(1987){van der Klis}, {Stella}, {White},
  {Jansen}, \& {Parmar}}]{1987ApJ...316..411V}
{van der Klis}, M., {Stella}, L., {White}, N., {Jansen}, F., \& {Parmar}, A.~N.
  1987, \apj, 316, 411, \dodoi{10.1086/165210}

\bibitem[{{Wang}(1981)}]{1981A&A...102...36W}
{Wang}, Y.~M. 1981, \aap, 102, 36

\bibitem[{{Wilms} {et~al.}(2000){Wilms}, {Allen}, \& {McCray}}]{Wilms2000}
{Wilms}, J., {Allen}, A., \& {McCray}, R. 2000, \apj, 542, 914,
  \dodoi{10.1086/317016}

\bibitem[{{Yuan} {et~al.}(2022){Yuan}, {Zhang}, {Chen}, \&
  {Ling}}]{2022hxga.book...86Y}
{Yuan}, W., {Zhang}, C., {Chen}, Y., \& {Ling}, Z. 2022, in Handbook of X-ray
  and Gamma-ray Astrophysics, 86, \dodoi{10.1007/978-981-16-4544-0_151-1}

\end{thebibliography}
\bibliographystyle{aasjournal}



\end{document}